\documentclass[superscriptaddress,onecolumn,pre]{revtex4}

\usepackage{amsmath}
\usepackage{graphicx,color}

\newcommand{\be}{\begin{equation}}
\newcommand{\ee}{\end{equation}}
\newcommand{\bea}{\begin{eqnarray}}
\newcommand{\eea}{\end{eqnarray}}

\begin{document}
\sloppy


\title{Kinetic theory of two-dimensional point vortices with collective effects}

\author{Pierre-Henri Chavanis}
\affiliation{Laboratoire de Physique Th\'eorique (IRSAMC), CNRS and UPS, Universit\'e de Toulouse, F-31062 Toulouse, France}

\begin{abstract}

We develop a kinetic theory of point vortices in two-dimensional
hydrodynamics taking collective effects into account. We first recall the approach
of Dubin \& O'Neil [Phys. Rev. Lett. {\bf 60}, 1286 (1988)] that leads
to a Lenard-Balescu-type kinetic equation for axisymmetric flows. When
collective effects are neglected, it reduces to the Landau-type
kinetic equation obtained independently in our previous papers [P.H. Chavanis,
Phys. Rev. E {\bf 64}, 026309 (2001); Physica A {\bf 387}, 1123 (2008)].
We also consider the relaxation
of a test vortex in a ``sea'' (bath) of field vortices. Its stochastic
motion is described in terms of a Fokker-Planck equation. We determine
the diffusion coefficient and the drift term by explicitly calculating
the first and second order moments of the radial displacement of the
test vortex from its equations of motion, taking collective effects
into account. This generalizes the expressions obtained in our
previous works. We discuss the scaling with $N$ of the relaxation time
for the system as a whole and for a test vortex in a bath.

\end{abstract}

\maketitle

\section{Introduction}
\label{sec_introduction}

The point vortex model \cite{newton} is interesting not only because
it provides a simplified model of two-dimensional (2D) hydrodynamics
with potential applications to large-scale geophysical and
astrophysical flows, but also because it constitutes a fundamental
example of systems with long-range interactions. Systems with
long-range interactions are numerous in nature (self-gravitating
systems, non-neutral plasmas, two-dimensional vortices,...) and they present
striking features that are absent in systems with short-range
interactions \cite{houches,assise,oxford}.  Their dynamics and
thermodynamics are actively studied at present \cite{cdr}.  The
statistical mechanics of the point vortex gas was first considered by
Onsager \cite{onsager} who discovered that negative temperature states
are possible for this system \footnote{His argument for the existence
of negative temperatures was given two years before Purcell \& Pound
\cite{purcell} reported the presence of negative ``spin temperatures''
in an experiment on nuclear spin systems.}. At negative temperatures (corresponding to high energies),
point vortices of the same sign have the tendency to cluster into
``supervortices'', similar to the large-scale vortices (e.g. Jupiter's
great red spot) observed in the atmosphere of giant planets. The
qualitative arguments of Onsager \cite{onsager} were developed more
quantitatively in a mean field approximation by Joyce \& Montgomery
\cite{jm,mj}, Kida \cite{kida} and Pointin \& Lundgren \cite{pl,lp}, and by Onsager himself in
unpublished notes \cite{esree}. The statistical theory predicts that the
point vortex gas should relax towards an equilibrium state described
by the Boltzmann distribution.  Specifically, the equilibrium stream
function is solution of a Boltzmann-Poisson equation. Many
mathematical works \cite{caglioti,k93,es,ca2,kl} have shown how a
proper thermodynamic limit can be rigorously defined for the point
vortex gas (in the Onsager picture). It is shown that the mean field
approximation becomes exact in the limit $N\rightarrow +\infty$ with
$\gamma\sim 1/N$ and $R\sim 1$ (where $N$ is the number of point
vortices, $\gamma$ is the circulation of a point vortex and $R$ is the system size).  However, the equilibrium statistical theory is based on some
assumptions of ergodicity and it is not proven that the point vortex
gas is ergodic \cite{khanin}. Therefore, it is not clearly established
that it will relax towards the Boltzmann distribution for
$t\rightarrow +\infty$. To settle this issue, and to determine the
relaxation time (in particular its scaling with $N$), we must develop
a kinetic theory of point vortices.

A kinetic theory has been developed by Dubin \& O'Neil \cite{dubin} in the context of non-neutral plasmas under a strong magnetic field, a system isomorphic  to the  point vortex
gas. They consider  axisymmetric mean flows and take
collective effects into account. Their approach is formal (but powerful) and uses the methods developed
in plasma physics. They perform a quasilinear theory of the Klimontovich equation, introduce Fourier-Laplace transforms and  work in the complex plane. They obtain a kinetic equation (\ref{lb27}) that is the counterpart of the Lenard-Balescu \cite{lenard,balescu} equation in plasma physics. Their work has been continued in \cite{sdprl,sd2,dubinjin,dubin2}, albeit in a  different perspective.

Independently, we have developed a kinetic theory of point vortices
\cite{preR,pre,bbgky,copenhaguen,kindetail} by using an analogy with
the kinetic theory of stellar systems
\cite{chandra,chandra1,nice,kandrup1,kandrup2}. The originality of our
approach is to remain in physical space so as to treat flows that are
not necessarily axisymmetric. We obtained a general kinetic equation
(\ref{lb29b}) that applies to arbitrary mean flows. We derived it by
different methods such as the projection operator formalism
\cite{pre}, the BBGKY hierarchy \cite{bbgky} and the quasilinear
theory \cite{bbgky}. Interestingly, the structure of this kinetic
equation bears a clear physical meaning in terms of generalized Kubo
relations. This equation is valid at the order $1/N$ and, for
$N\rightarrow +\infty$, it reduces to the (smooth) 2D Euler equation
(\ref{pvm13}) which describes the collisionless evolution of the point
vortex gas. For axisymmetric mean flows, we obtained a simpler and more
explicit kinetic equation (\ref{lb28}) that is the counterpart of the
Landau \cite{landau} equation in plasma physics. Our formalism
\cite{bbgky,kindetail} is simple and clearly shows the structure of
the kinetic equation and the physical origin of its different
components. Furthermore, it avoids the use of Fourier-Laplace
transforms and provides a direct and transparent derivation of the
kinetic equation without going into the complex plane (which, somehow,
hides the basic physics). The derived kinetic equation conserves the
circulation, the energy and the angular momentum. It also increases
monotonically the Boltzmann entropy ($H$-theorem). The collisional
evolution is due to a condition of resonance between distant vortices
and the relaxation stops when the profile of angular velocity becomes
monotonic, even if the system has not reached the Boltzmann
distribution of statistical equilibrium \cite{clemou}.  Since the
kinetic equation is valid at the order $1/N$, this ``kinetic
blocking'' implies that the relaxation time for an axisymmetric
distribution of point vortices is larger than $Nt_{D}$, where $t_D$ is
the dynamical time. In fact, there is no proof so far from the kinetic
theory that the system relaxes towards the Boltzmann statistical
equilibrium for $t\rightarrow +\infty$. To settle this issue, we need
to develop the kinetic theory at higher orders (taking into account
more complex three-body, four-body... correlations between vortices).
By contrast, for a non-axisymmetric evolution, there are potentially more
resonances and the relaxation time can be shorter, of the order of the
natural timescale $Nt_D$
\cite{kindetail}.
We also considered the stochastic motion of a test vortex in a ``sea''
(bath) of field vortices
\cite{preR,pre,bbgky,copenhaguen,kindetail}. It can be described in
terms of a Fokker-Planck equation involving a diffusion term and a
drift term. For a thermal bath, they are connected to each other by an
Einstein relation. The diffusion coefficient (resp. drift term) is
proportional to the vorticity distribution (resp. to the gradient of
the vorticity distribution) of the field vortices and inversely
proportional to the local shear, a feature first noted in \cite{preR}.
The distribution of the test vortex relaxes towards the distribution
of the field vortices (bath) on a timescale $(N/\ln N)t_{D}$.

A formal limitation of our approach is that it neglects collective
effects. This is not so crucial as long as we are just interested in
the main properties of the kinetic equation that do not sensibly
depend on collective effects (as we shall see, collective effects
essentially amount to replacing the ``bare'' potential of interaction
by a ``dressed'' potential of interaction, without altering the
overall structure of the kinetic equation). However, collective
effects may change the results quantitatively \footnote{Recently, Sano
\cite{sano} confirmed our generalized Landau-type kinetic equation
(\ref{lb29b}) by using a BBGKY approach similar to the one developed
in \cite{bbgky}. He also proposed a Lenard-Balescu-type equation,
taking into account collective effects, by using the theory of the
Fredholm integral equation. However, he argues that this equation
reduces to the Landau-type equation for large $N$ although our present
work shows that both equations (Landau and Lenard-Balescu-type) are
valid at the same order $1/N$. The difference is that the Landau-type
equation is an approximation of the Lenard-Balescu-type equation since
it ignores collective effects.}.  In this paper, we generalize our
kinetic theory so as to take collective effects into account. We
also extend the kinetic theory to a general class of potentials of
interaction between point vortices (in particular, they may include
screening effects that are important in geophysics
\cite{pedlosky}). In Sec. \ref{sec_whole}, we consider the evolution
of the system as a whole and re-derive the Lenard-Balescu-type kinetic
equation obtained in Ref. \cite{dubin}. This is done for the sake of
completeness and in order to obtain intermediate results that will be
useful in the sequel.  We also show the connection with the
Landau-type kinetic equation that we independently derived in Refs.
\cite{preR,pre,bbgky,copenhaguen,kindetail} with another method when
collective effects are neglected. In Sec. \ref{sec_stoch}, we consider
the relaxation of a test vortex in a bath of field vortices and derive
a Fokker-Planck equation (this Fokker-Planck approach is not
considered in \cite{dubin}). We determine the diffusion coefficient
and the drift term by explicitly calculating the first and second
order moments of the radial displacement of the test vortex from its
equations of motion, taking collective effects into account. This
generalizes the expressions obtained in our previous works
\cite{preR,pre,bbgky,copenhaguen,kindetail} where collective effects are neglected. Finally, we show the connection
between the Fokker-Planck equation describing the relaxation of a test
vortex in a bath and the kinetic equation describing the evolution of
the system as a whole and compare the relaxation time in these two
situations.

There exists many analogies between systems with long-range interactions although they are of a very different nature. We have already mentioned analogies between plasmas, stellar systems and point vortices \footnote{The analogy between the statistical mechanics and the kinetic theory of stellar systems and two-dimensional vortices is developed in \cite{csr,houchesPH,kindetail}.}. There also exists analogies with a toy model of systems with long-range interactions called the Hamiltonian Mean Field (HMF) model \cite{ar}. The kinetic theory of this model is developed in \cite{inagakikin,bouchet,bd,cvb,clpre,curious,cdr,bgm,kindetail} and it presents features that are similar to those observed earlier for point vortices \cite{dubin,preR,pre}. In particular, the Lenard-Balescu collision term vanishes in one dimension due to the absence of resonances (a result known long ago in plasma physics \cite{feix,kp} and rediscovered in the context of the HMF model \cite{bd,cvb}) so that the relaxation time of the spatially homogeneous  HMF model is larger than $Nt_D$ (it can be reduced to $Nt_D$ if the system is spatially inhomogeneous due to additional resonances \cite{angleaction,kindetail}). On the other hand, the relaxation of a test particle in a bath is described by a Fokker-Planck equation involving a term of diffusion and a term of friction. This is similar to the Fokker-Planck equation derived by Chandrasekhar \cite{chandra1,nice} in stellar dynamics and by Chavanis \cite{preR,pre,bbgky,copenhaguen,kindetail} for the point vortex system. This is also  a particular case (in one dimension and for a potential of interaction truncated to one Fourier mode) of the general Fokker-Planck equation  derived by Hubbard \cite{hubbard} in plasma physics. In order to stress the analogies between these different topics, we shall use a presentation that is similar to the one given in the review of Campa {\it et al.} \cite{cdr} on long-range interacting systems.

\section{The point vortex model}
\label{sec_pvm}

Let us consider a system of point vortices with individual circulation $\gamma$ moving on the infinite plane. Their dynamics is described by the Hamilton-Kirchhoff \cite{kirchhoff,newton} equations
\begin{eqnarray}
\gamma \frac{dx_i}{dt}=\frac{\partial H}{\partial y_i},\qquad \gamma \frac{dy_i}{dt}=-\frac{\partial H}{\partial x_i},
\label{pvm1}
\end{eqnarray}
with the Hamiltonian
\begin{eqnarray}
H=\gamma^2\sum_{i<j}u(|{\bf r}_i-{\bf r}_j|).
\label{pvm2}
\end{eqnarray}
We note that the coordinates $x$ and $y$ of the point vortices are canonically conjugate. The standard potential of interaction between point vortices is solution of the Poisson equation $\Delta u=-\delta$. In an infinite domain, it is given by $u(|{\bf r}-{\bf r}'|)=-(1/2\pi)\ln |{\bf r}-{\bf r}'|$. However, we shall let the function $u$ arbitrary in order to treat more general situations. For example, in the quasigeostrophic (QG) approximation of geophysical fluid dynamics, the Poisson equation is replaced by the screened Poisson equation $\Delta u-k_R^2 u=-\delta$ where $k_R^{-1}$ is the Rossby radius \cite{pedlosky}. Equation (\ref{pvm1}) can be rewritten in vectorial form
\begin{eqnarray}
\gamma \frac{d{\bf r}_i}{dt}=-{\bf z}\times\nabla_i H,
\label{pvm3}
\end{eqnarray}
where ${\bf z}$ is a unit vector normal to the plane on which the vortices move. In an infinite domain, the point vortex system conserves the total circulation $\Gamma=N\gamma$, the energy $E=H$ and the angular momentum $L=\sum_i \gamma r_i^2$ (the center of ``mass'' ${\bf R}=\sum_i \gamma {\bf r}_i$ is also conserved and taken as the origin $O$ of the system of coordinates). We introduce the discrete vorticity field
\begin{eqnarray}
\omega_d({\bf r},t)=\gamma\sum_i\delta({\bf r}-{\bf r}_i(t)).
\label{pvm4}
\end{eqnarray}
Differentiating this expression with respect to time and using the equation of motion (\ref{pvm3}), we get
\begin{eqnarray}
\frac{\partial\omega_d}{\partial t}=\sum_i ({\bf z}\times \nabla_i H)\cdot \nabla \delta({\bf r}-{\bf r}_i(t)).
\label{pvm5}
\end{eqnarray}
This equation can be rewritten
\begin{eqnarray}
\frac{\partial\omega_d}{\partial t}=\nabla \cdot \sum_i ({\bf z}\times \nabla_i H) \delta({\bf r}-{\bf r}_i(t))=\gamma\nabla \cdot \sum_i ({\bf z}\times \nabla\psi_d) \delta({\bf r}-{\bf r}_i(t)),
\label{pvm6}
\end{eqnarray}
where
\begin{eqnarray}
\psi_d({\bf r},t)=\int u(|{\bf r}-{\bf r}'|)\omega_d({\bf r}',t)\, d{\bf r}',
\label{pvm7}
\end{eqnarray}
is the discrete stream function induced by the discrete vorticity field (we have used the $\delta$-function to replace $\nabla\psi_d({\bf r}_i(t),t)$ by $\nabla\psi_d({\bf r},t)$ in the last equality of Eq. (\ref{pvm6})). Introducing the discrete velocity field ${\bf u}_d=-{\bf z}\times \nabla\psi_d$, we obtain
\begin{eqnarray}
\frac{\partial\omega_d}{\partial t}+\nabla (\omega_d {\bf u}_d)=0.
\label{pvm8}
\end{eqnarray}
Finally, using the incompressibility condition $\nabla\cdot {\bf u}_d=0$, we can rewrite this equation in the form
\begin{eqnarray}
\frac{\partial\omega_d}{\partial t}+{\bf u}_d\cdot \nabla\omega_d=0.
\label{pvm9}
\end{eqnarray}
This equation is {\it exact} and contains the same information as the Hamiltonian system (\ref{pvm1})-(\ref{pvm2}). This is the counterpart of the Klimontovich equation in plasma physics \cite{pitaevskii}. This is also the 2D Euler equation from which the point vortex model is issued \cite{newton}.

We now introduce a smooth vorticity field $\omega({\bf r},t)=\langle \omega_{d}({\bf r},t)\rangle$ corresponding to an average of $\omega_{d}({\bf r},t)$ over a large number of initial conditions. We then write $\omega_d=\omega+\delta\omega$ where $\delta\omega$ denotes the fluctuations around the smooth distribution. Similarly, we write $\psi_d=\psi+\delta\psi$ and ${\bf u}_d={\bf u}+\delta {\bf u}$. Substituting these decompositions in Eq. (\ref{pvm9}), we obtain
\begin{equation}
\frac{\partial\omega}{\partial t}+\frac{\partial\delta\omega}{\partial t}+{\bf u}\cdot\nabla\omega+{\bf u}\cdot\nabla\delta\omega+\delta{\bf u}\cdot\nabla\omega+\delta{\bf u}\cdot\nabla\delta\omega=0.
\label{pvm10}
\end{equation}
Taking the average of this equation over the initial conditions, we get
\begin{equation}
\frac{\partial\omega}{\partial t}+{\bf u}\cdot\nabla\omega=-\nabla\cdot \langle \delta\omega \delta{\bf u}\rangle,
\label{pvm11}
\end{equation}
where the right hand side can be interpreted as a ``collision'' term. Of course, it does not correspond to direct collisions between vortices but to distant collisions, or correlations, due to the long-range interaction.
Subtracting this expression from Eq. (\ref{pvm10}), we obtain
\begin{equation}
\frac{\partial\delta\omega}{\partial t}+{\bf u}\cdot\nabla\delta\omega+\delta{\bf u}\cdot\nabla\omega=\nabla\cdot \langle \delta\omega \delta{\bf u}\rangle-\nabla\cdot (\delta\omega \delta{\bf u}).
\label{pvm12}
\end{equation}
These equations are still exact. We now consider the thermodynamic limit $N\rightarrow +\infty$ with $\gamma\sim 1/N$. In this limit, we have $\delta\omega\sim 1/\sqrt{N}$ and $\delta\psi\sim 1/\sqrt{N}$ and  we see that the collision term scales like $1/N$. Therefore, for $N\rightarrow +\infty$, Eq. (\ref{pvm11}) reduces to the (smooth) 2D Euler equation
\begin{equation}
\frac{\partial\omega}{\partial t}+{\bf u}\cdot\nabla\omega=0,\qquad {\bf u}=-{\bf z}\times \nabla\psi,\qquad \psi({\bf r},t)=\int u(|{\bf r}-{\bf r}'|)\omega({\bf r}',t)\, d{\bf r}'.
\label{pvm13}
\end{equation}
The 2D Euler equation is obtained when ``collisions'' (more properly
granular effects, discreteness effects or finite $N$ corrections)
between point vortices are neglected. The 2D Euler equation can also
be obtained from the BBGKY hierarchy by neglecting correlations
between point vortices \cite{bbgky}. This is valid, for a fixed
interval of time, when $N \rightarrow +\infty$. In that case, the mean
field approximation becomes {\it exact} and the $N$-body distribution
function can be factorized as a product of $N$ one-body distributions,
resulting in Eq. (\ref{pvm13}). The 2D Euler equation (\ref{pvm13}) is
the counterpart of the Vlasov equation in plasma physics
\cite{pitaevskii} and stellar dynamics \cite{bt}. Starting from an
unsteady or unstable initial condition, the 2D Euler equation is known
to develop a complicated mixing process leading to the formation of a
quasi stationary state (QSS) on a coarse-grained scale
\cite{chen,brands}. This takes place on a very short timescale, of the
order of the dynamical time $t_D\sim \omega^{-1}$, where $\omega\sim
\Gamma/R^2$ is the typical vorticity ($R$ is the system size
\footnote{When the domain is infinite, the vorticity region
extends over a typical distance $R$ which will be identified with the
system size.}). This process, which is purely collisionless and driven by
mean field effects, is known as violent relaxation. A statistical
theory of the 2D Euler equation has been developed by Miller
\cite{miller} and Robert \& Sommeria \cite{rs} to predict the QSS that
results from violent relaxation assuming ergodicity. This is the
counterpart of the statistical theory of the Vlasov equation
introduced by Lynden-Bell \cite{lb} in astrophysics to describe the
violent relaxation of collisionless stellar systems. The process of
violent relaxation towards a non-Boltzmannian QSS has also been
extensively studied for the HMF model \cite{cdr}. In some cases,
violent relaxation is {\it incomplete} due to lack of efficient mixing
\footnote{Numerical simulations have been made in the
``two levels'' case where the vorticity $\omega$ (resp. distribution
function $f$) takes only two values $\sigma_0$ and $0$ (resp. $\eta_0$
and $0$) \cite{brands,chen,joyceLB}. Generically, one observes the
existence of a ``mixing zone'' characterized by a linear relationship
between $\ln(\overline{\omega}/(\sigma_0-\overline{\omega}))$ and
$\psi_*$ (resp. $\ln(\overline{f}/(\eta_0-\overline{f}))$ and
$\epsilon$). This corresponds to a {\it local} Lynden-Bell
distribution $\overline{\omega}=\sigma_0/(1+{\rm
exp}(\beta'\psi_*+\alpha'))$ (resp. $\overline{f}=\eta_0/(1+{\rm
exp}(\beta'\epsilon+\alpha'))$) with parameters $\beta'$ and $\alpha'$
different from the global ones $\beta$ and $\alpha$. On the other
hand, violent relaxation is incomplete in the core and in the halo
resulting in deviations from this linear relationship for high and low
$\overline{\omega}$ (resp. $\overline{f}$). In particular, the peak
vorticity (resp. distribution function) has the tendency to be
conserved and the system has the tendency to remain confined. This can
be explained by a kinetic theory of collisionless violent relaxation
\cite{bbgky,hb4}. The evolution of $\overline{\omega}$
(resp. $\overline{f}$) is governed by a kinetic equation of the Landau
type with a term describing relaxation towards the Lynden-Bell
distribution. One feature of this term is that the relaxation current
involves the product $\overline{\omega}(\sigma_0-\overline{\omega})$
(resp. $\overline{f}(\eta_0-\overline{f})$). This implies that
relaxation is {\it slow} in the core ($\overline{\omega}\rightarrow
\sigma_0$, $\overline{f}\rightarrow \eta_0$) and in the tail
($\overline{\omega}\rightarrow 0$, $\overline{f}\rightarrow
0$). Since, in addition, the efficiency of relaxation decreases with
time (through a prefactor $\epsilon(t)\rightarrow 0$ related to the
correlation length in the kinetic equation), the system will not have
time to mix well in these regions. Therefore, the vorticity
(resp. distribution function) will be higher in the core and lower in
the tail than predicted by Lynden-Bell's statistical theory. In the
intermediate region, on the contrary, the system will mix efficiently
and the kinetic equation will rapidly converge towards a local
Lynden-Bell distribution. This theoretically motivated scenario
\cite{bbgky,hb4} seems to be consistent with  the
numerical simulations of 2D turbulence (see Fig. 2 in
\cite{chen} and Fig. 5 in
\cite{brands}) and 1D self-gravitating systems (see Fig. 20 in
\cite{joyceLB}). Note that in case of incomplete relaxation (non-ergodicity), the QSS is sometimes found to be close to a polytropic (Tsallis) distribution \cite{boghosian,brands,chavcampa} but this is not general.}.

On longer timescales, ``collisions'' (more properly correlations) between point vortices develop and the system deviates from the 2D Euler dynamics. For $t\rightarrow +\infty$, we expect that the system will reach a statistical equilibrium state. In the thermodynamic limit $N \rightarrow +\infty$ with $\gamma\sim 1/N$, the mean field approximation is exact \cite{caglioti,k93,es,ca2,kl}. The statistical equilibrium state in the microcanonical ensemble is obtained by maximizing the Boltzmann entropy $S=-\int (\omega/\gamma)\ln(\omega/\gamma)\, d{\bf r}$ at fixed circulation $\Gamma=\int\omega\, d{\bf r}$, energy $E=(1/2)\int\omega\psi\, d{\bf r}$ and angular momentum $L=\int\omega r^2\, d{\bf r}$ \cite{jm,clemou}. This variational principle determines the most probable macrostate $\omega({\bf r})$. Writing the first order variations as $\delta S-\beta\delta E-\alpha\delta \Gamma-(1/2)\beta\Omega_L \delta L=0$ where $\beta$, $\alpha$ and $\Omega_L$ are Lagrange multipliers, we obtain the Boltzmann distribution
\begin{equation}
\omega=Ae^{-\beta\gamma\psi_{*}},\qquad \psi({\bf r})=\int u(|{\bf r}-{\bf r}'|)\omega({\bf r}')\, d{\bf r}',
\label{pvm14}
\end{equation}
where $\psi_{*}=\psi+(1/2)\Omega_L r^2$ is the relative stream function taking into account the invariance by rotation of the system. For the ordinary interaction between point vortices, the equilibrium stream function is solution of the Boltzmann-Poisson equation \footnote{Equation (\ref{pvm14}), or the Boltzmann-Poisson equation, determines all the critical points of constrained entropy, i.e. those that cancel the first order variations. Of course, among these solutions, only entropy {\it maxima} are physically relevant. We must therefore consider the sign of the second order variations of entropy.}. We must remember, however, that the statistical theory is based on an assumption of ergodicity and on the postulate that all the accessible microstates are equiprobable. There is no guarantee that this is true for the point vortex model \cite{khanin}. If we want to prove that the point vortex gas truly reaches the Boltzmann distribution (\ref{pvm14}), and if we want to determine the relaxation time (in particular its scaling with the number $N$ of point vortices), we must develop a kinetic theory of point vortices.

\section{Collisional evolution of the system as a whole}
\label{sec_whole}

\subsection{Lenard-Balescu and Landau-type kinetic equations}
\label{sec_lb}

As we have previously indicated, the ``collisions'' between point vortices can be neglected for times shorter than $N t_D$ where $t_D$ is the dynamical time. If we want to describe the  evolution of the system on a longer time scale, we must take finite $N$ corrections into account. At the level $1/N$, we can neglect the quadratic term on the right hand side of Eq. (\ref{pvm12}). Indeed, the left hand side is of order $1/\sqrt{N}$ and the right hand side is of order $1/N\ll 1/\sqrt{N}$. We therefore obtain the set of coupled equations
\begin{equation}
\frac{\partial\omega}{\partial t}+{\bf u}\cdot\nabla\omega=-\nabla\cdot \langle \delta\omega \delta{\bf u}\rangle,
\label{lb1}
\end{equation}
\begin{equation}
\frac{\partial\delta\omega}{\partial t}+{\bf u}\cdot\nabla\delta\omega+\delta{\bf u}\cdot\nabla\omega=0.
\label{lb2}
\end{equation}
They form the starting point of the quasilinear theory.  If we restrict ourselves to axisymmetric mean flows, we can write  ${\bf u}({\bf r},t)=u(r,t){\bf e}_{\theta}$ with $u(r,t)=-\partial\psi/\partial r(r,t)=\Omega(r,t)r$, where  $\Omega(r,t)=r^{-2}\int_0^r \omega(r',t)r'\, dr'$  is the angular velocity field (see Sec. 6.1 of \cite{clemou}). In that case, Eqs. (\ref{lb1})-(\ref{lb2}) reduce to the coupled equations
\begin{equation}
\frac{\partial\omega}{\partial t}=-\frac{1}{r}\frac{\partial}{\partial r}\left \langle \delta\omega \frac{\partial\delta\psi}{\partial\theta}\right \rangle,
\label{lb3}
\end{equation}
\begin{equation}
\frac{\partial\delta\omega}{\partial t}+\Omega\frac{\partial\delta\omega}{\partial\theta}+\frac{1}{r}\frac{\partial\delta\psi}{\partial \theta}\frac{\partial\omega}{\partial r}=0.
\label{lb4}
\end{equation}
We shall assume that the fluctuations evolve rapidly compared to the transport time scale, so that time variation of $\omega$, $\psi$ and $\Omega$ can be neglected in the calculation of the collision term (Bogoliubov ansatz). For brevity, we shall omit the variable $t$ in these quantities. We also assume that the axisymmetric distribution always remain Euler stable, so that it evolves only under the effect of collisions and not because of dynamical instabilities.  In that case, Eqs. (\ref{lb3})-(\ref{lb4}) can be solved with the aid of Fourier-Laplace transforms and the collision term can be explicitly calculated. This is the kinetic approach  developed by Dubin \& O'Neil \cite{dubin}. We shall recall the main steps of their analysis by using a close parallel with the derivation of the Lenard-Balescu equation in plasma physics \cite{pitaevskii,paddy,hb2,cdr}. This will allow us to set the notations and derive useful intermediate results that will be needed in the sequel.

The Fourier-Laplace transform of the fluctuations of the vorticity field $\delta\omega$ is defined by
\begin{equation}
\delta \tilde\omega(n,r,\sigma)=\int_{0}^{2\pi}\frac{d\theta}{2\pi}\int_{0}^{+\infty}dt\, e^{-i(n\theta-\sigma t)}\delta\omega(\theta,r,t).
\label{lb5}
\end{equation}
This expression for the Laplace transform is valid for ${\rm Im}(\sigma)$ sufficiently large. For the remaining part of the complex $\sigma$ plane, it is defined by an analytic continuation. The inverse transform is
\begin{equation}
\delta \omega(\theta,r,t)=\sum_{n=-\infty}^{+\infty}\int_{\cal C}\frac{d\sigma}{2\pi}\, e^{i(n\theta-\sigma t)}\delta\tilde\omega(n,r,\sigma),
\label{lb6}
\end{equation}
where the Laplace contour ${\cal C}$ in the complex $\sigma$ plane must pass above all poles of the integrand. Similar expressions hold for the fluctuations of the stream function $\delta\psi$. If we take the  Fourier-Laplace transform of Eq. (\ref{lb4}), we find that
\begin{equation}
-\delta\hat{\omega}(n,r,0)-i\sigma\delta\tilde\omega(n,r,\sigma)+in\Omega\delta\tilde\omega(n,r,\sigma)+in\frac{1}{r}\frac{\partial\omega}{\partial r}\delta\tilde\psi(n,r,\sigma)=0,
\label{lb7}
\end{equation}
where the first term is the spatial Fourier transform of the initial value
\begin{equation}
\delta\hat\omega(n,r,0)=\int_{0}^{2\pi}\frac{d\theta}{2\pi}\, e^{-in\theta}\delta\omega(\theta,r,0).
\label{lb8}
\end{equation}
The foregoing equation can be rewritten
\begin{equation}
\delta\tilde\omega (n,r,\sigma)=-\frac{n\frac{1}{r}\frac{\partial\omega}{\partial r}}{n\Omega-\sigma}\delta\tilde\psi(n,r,\sigma)+\frac{\delta\hat\omega(n,r,0)}{i(n\Omega-\sigma)},
\label{lb9}
\end{equation}
where the first term on the right hand side corresponds to ``collective effects'' and the second term is related to the initial condition. The fluctuations of the stream function are related to the fluctuations of the vorticity by an equation of the form $L\delta\psi=-\delta\omega$ equivalent to $\delta\psi=u*\delta\omega$ (see Appendix \ref{sec_pot}).  Taking the Fourier-Laplace transform of this equation, we obtain ${\cal L}\delta\tilde\psi(n,r,\sigma)=-\delta\tilde\omega(n,r,\sigma)$. Using Eq. (\ref{lb9}), it can be rewritten in the form \footnote{This equation with the right hand side equal to zero is the Rayleigh equation determining the dispersion relation associated with the linearized 2D Euler equation.}
\begin{equation}
\left\lbrack {\cal L}-\frac{n\frac{1}{r}\frac{\partial\omega}{\partial r}}{n\Omega-\sigma}\right \rbrack \delta\tilde\psi(n,r,\sigma)=-\frac{\delta\hat\omega(n,r,0)}{i(n\Omega-\sigma)}.
\label{lb10}
\end{equation}
Therefore, the Fourier-Laplace transform of the fluctuations of the stream function is related to the initial condition by
\begin{equation}
\delta\tilde\psi(n,r,\sigma)=\int_{0}^{+\infty}2\pi r'\, dr'\, G(n,r,r',\sigma)\frac{\delta\hat\omega(n,r',0)}{i(n\Omega'-\sigma)},
\label{lb11}
\end{equation}
where the Green function $G(n,r,r',\sigma)$ is defined by
\begin{equation}
\left\lbrack {\cal L}-\frac{n\frac{1}{r}\frac{\partial\omega}{\partial r}}{n\Omega-\sigma}\right \rbrack G(n,r,r',\sigma)=-\frac{\delta(r-r')}{2\pi r}.
\label{lb12}
\end{equation}
The Fourier-Laplace transform of the fluctuations of the vorticity is
then given by Eq. (\ref{lb9}) with Eq. (\ref{lb11}). In these equations,  we have noted
$\Omega$ for $\Omega(r)$ and $\Omega'$ for $\Omega(r')$. Similarly, we shall note
$\omega$ for $\omega(r)$ and $\omega'$ for $\omega(r')$. To avoid confusion, the derivatives with respect to $r$ will be noted $\partial/\partial r$.

We can use these expressions to compute the collisional term appearing on the right hand side of Eq. (\ref{lb3}). One has
\begin{equation}
\left\langle\delta\omega\frac{\partial\delta\psi}{\partial\theta}\right\rangle=\sum_{n}\sum_{n'}\int_{\cal C}\frac{d\sigma}{2\pi}\int_{\cal C}\frac{d\sigma'}{2\pi} \, i n' e^{i(n\theta-\sigma t)}e^{i(n'\theta-\sigma' t)}\langle \delta\tilde\omega(n,r,\sigma)\delta\tilde\psi(n',r,\sigma')\rangle.
\label{lb13}
\end{equation}
Using Eq. (\ref{lb9}), we find that
\begin{equation}
\langle \delta\tilde\omega(n,r,\sigma)\delta\tilde\psi(n',r,\sigma')\rangle=-\frac{n\frac{1}{r}\frac{\partial\omega}{\partial r}}{n\Omega-\sigma}\langle \delta\tilde\psi(n,r,\sigma)\delta\tilde\psi(n',r,\sigma')\rangle+\frac{\langle \delta\hat\omega(n,r,0)\delta\tilde\psi(n',r,\sigma')\rangle}{i(n\Omega-\sigma)}.
\label{lb14}
\end{equation}
The first term corresponds to the self-correlation of the stream function, while the second term corresponds to the correlations between the fluctuations of the stream function and of the vorticity at time $t=0$. Let us consider these two terms separately.

From Eq. (\ref{lb11}), we obtain
\begin{equation}
\langle \delta\tilde\psi(n,r,\sigma)\delta\tilde\psi(n',r,\sigma')\rangle=-\int_{0}^{+\infty}2\pi r'\, dr'\int_{0}^{+\infty}2\pi r''\, dr'' \, G(n,r,r',\sigma)G(n',r,r'',\sigma')\frac{\langle \delta\hat\omega(n,r',0)\delta\hat\omega(n',r'',0)\rangle}{(n\Omega'-\sigma)(n'\Omega''-\sigma')}.
\label{lb15}
\end{equation}
Using the expression of the auto-correlation of the fluctuations at $t=0$ (see Appendix \ref{sec_auto}) given by
\begin{equation}
\langle \delta\hat\omega(n,r',0)\delta\hat\omega(n',r'',0)\rangle={\gamma}\delta_{n,-n'}\frac{\delta(r'-r'')}{2\pi r'}\omega(r'),
\label{lb16}
\end{equation}
we find that
\begin{equation}
\langle \delta\tilde\psi(n,r,\sigma)\delta\tilde\psi(n',r,\sigma')\rangle=2\pi\gamma \, \delta_{n,-n'} \int_{0}^{+\infty} r'\, dr' \, G(n,r,r',\sigma)G(-n,r,r',\sigma')\frac{\omega(r')}{(n\Omega'-\sigma)(n\Omega'+\sigma')}.
\label{lb17}
\end{equation}
Considering only the contributions that do not decay in time, it can be shown \cite{pitaevskii,cdr} that $\lbrack (n\Omega'-\sigma)(n\Omega'+\sigma')\rbrack^{-1}$ can be substituted by $(2\pi)^2\delta(\sigma+\sigma')\delta(\sigma-n\Omega')$. Then, using the property $G(-n,r,r',-\sigma)=G(n,r,r',\sigma)^*$ \cite{dubin}, one finds that the correlations of the fluctuations of the stream function are given by
\begin{equation}
\langle \delta\tilde\psi(n,r,\sigma)\delta\tilde\psi(n',r,\sigma')\rangle=(2\pi)^3 \gamma \, \delta_{n,-n'} \delta(\sigma+\sigma')\int_{0}^{+\infty} r'\, dr' \,  |G(n,r,r',\sigma)|^2 \delta(\sigma-n\Omega')\omega(r').
\label{lb18}
\end{equation}
This is an important relation that we wish to emphasize.
Similarly, one finds that the second term on the right hand side of Eq. (\ref{lb14}) is given by
\begin{equation}
\frac{\langle \delta\hat\omega(n,r,0)\delta\tilde\psi(n',r,\sigma')\rangle}{i(n\Omega-\sigma)}=(2\pi)^2 \gamma \, \delta_{n,-n'} \delta(\sigma+\sigma') G(-n,r,r,-\sigma) \delta(\sigma-n\Omega)\omega(r).
\label{lb19}
\end{equation}

From Eq. (\ref{lb18}), we get the contribution to (\ref{lb13}) of the first term of Eq. (\ref{lb14}). It is given by
\begin{equation}
\left\langle\delta\omega\frac{\partial\delta\psi}{\partial\theta}\right\rangle_{I}=i (2\pi)^2 \gamma \sum_{n}\int_{\cal C}\frac{d\sigma}{2\pi}\int_{0}^{+\infty} r'\, dr'  \,  \frac{n^2\frac{1}{r}\frac{\partial\omega}{\partial r}}{n\Omega-\sigma} |G(n,r,r',\sigma)|^2 \delta(\sigma-n\Omega') \omega(r').
\label{lb20}
\end{equation}
Using the Plemelj formula,
\begin{equation}
\lim_{\epsilon\rightarrow 0}\frac{1}{x-a\pm i 0^+}={\cal P}\frac{1}{x-a}\mp i\pi\delta(x-a),
\label{lb21}
\end{equation}
where ${\cal P}$ denotes the principal value, integrating over $\sigma$ and using the identity
\begin{equation}
\delta(\lambda x)=\frac{1}{|\lambda|}\delta(x),
\label{lb22}
\end{equation}
we obtain
\begin{equation}
\left\langle\delta\omega\frac{\partial\delta\psi}{\partial\theta}\right\rangle_{I}=-2\pi^2 \gamma \sum_{n}\int_{0}^{+\infty} r'\, dr'  \, |n| |G(n,r,r',n\Omega)|^2 \delta(\Omega-\Omega')\omega(r') \frac{1}{r}\frac{\partial\omega}{\partial r}.
\label{lb23}
\end{equation}

From Eq. (\ref{lb19}), we get the contribution to (\ref{lb13}) of the second term of Eq. (\ref{lb14}). It is given by
\begin{equation}
\left\langle\delta\omega\frac{\partial\delta\psi}{\partial\theta}\right\rangle_{II}=2\pi \gamma \sum_{n}\int_{\cal C}\frac{d\sigma}{2\pi}\, n \, {\rm Im}\, G(n,r,r,\sigma)\delta(\sigma-n\Omega) \omega(r).
\label{lb24}
\end{equation}
To determine ${\rm Im}\, G(n,r,r,\sigma)$, we multiply Eq. (\ref{lb12}) by $2\pi r\, G(-n,r,r',-\sigma)$, integrate over $r$ from $0$ to $+\infty$ and take the imaginary part of the resulting expression using the Plemelj formula. This yields
\begin{equation}
{\rm Im}\,  G(n,r,r,\sigma)=\mp \, 2\pi^2\int_{0}^{+\infty} dr'  \, n |G(n,r,r',\sigma)|^2 \delta(\sigma-n\Omega')\frac{\partial\omega'}{\partial r'}.
\label{lb25}
\end{equation}
Substituting this expression in Eq. (\ref{lb24}) and using Eq. (\ref{lb22}), we obtain
\begin{equation}
\left\langle\delta\omega\frac{\partial\delta\psi}{\partial\theta}\right\rangle_{II}= 2\pi^2 \gamma \sum_{n}\int_{0}^{+\infty} r'\, dr'  \, |n| |G(n,r,r',n\Omega)|^2 \delta(\Omega-\Omega')\omega \frac{1}{r'}\frac{\partial\omega'}{\partial r'}.
\label{lb26}
\end{equation}
Finally, regrouping Eqs. (\ref{lb3}), (\ref{lb23}) and (\ref{lb26}), we end up on the kinetic equation
\begin{equation}
\frac{\partial\omega}{\partial t}= 2\pi^2 \gamma \frac{1}{r}\frac{\partial}{\partial r}\sum_{n}\int_{0}^{+\infty} r'\, dr'  \, |n| |G(n,r,r',n\Omega)|^2 \delta(\Omega-\Omega')\left (\frac{1}{r}\omega'\frac{\partial\omega}{\partial r}-\frac{1}{r'}\omega\frac{\partial\omega'}{\partial r'}\right ).
\label{lb27}
\end{equation}
This equation, taking  collective effects into account, was derived by Dubin \& O'Neil \cite{dubin}. It is the counterpart of the Lenard-Balescu equation in plasma physics. When collective effects are neglected, we independently obtained in Refs. \cite{pre,bbgky,copenhaguen,kindetail} a kinetic equation of the form
\begin{equation}
\frac{\partial\omega}{\partial t}= 2\pi^2 \gamma \frac{1}{r}\frac{\partial}{\partial r}\sum_{n}\int_{0}^{+\infty} r'\, dr'  \, |n| \hat{u}_n(r,r')^2 \delta(\Omega-\Omega')\left (\frac{1}{r}\omega'\frac{\partial\omega}{\partial r}-\frac{1}{r'}\omega\frac{\partial\omega'}{\partial r'}\right ).
\label{lb28}
\end{equation}
For the usual potential of interaction, the sum over $n$ can be done explicitly (see Appendix \ref{sec_pot}) and we get
\begin{equation}
\frac{\partial\omega}{\partial t}= -\frac{\gamma}{4} \frac{1}{r}\frac{\partial}{\partial r}\int_{0}^{+\infty} r'\, dr'  \, \ln\left\lbrack 1-\left (\frac{r_<}{r_>}\right )^2\right \rbrack \delta(\Omega-\Omega')\left (\frac{1}{r}\omega'\frac{\partial\omega}{\partial r}-\frac{1}{r'}\omega\frac{\partial\omega'}{\partial r'}\right ),
\label{lb29}
\end{equation}
where $r_<$ (resp. $r_>$) is the min (resp. max) of $r$ and $r'$. This kinetic equation was derived with a  method that does not use Fourier-Laplace transforms. It arises from the generalized kinetic equation
\begin{eqnarray}
\frac{\partial \omega}{\partial t}+\frac{N-1}{N} {\bf
u}\cdot \nabla\omega
=\frac{\partial}{\partial {r}^{\mu}}\int_0^{t} d\tau \int d{\bf
r}_{1} {V}^{\mu}(1\rightarrow 0){\cal G}(t,t-\tau)\nonumber\\
\times  \left \lbrack {{\cal V}}^{\nu}(1\rightarrow 0)
{\partial\over\partial {r}^{\nu}}+{{\cal V}}^{\nu}(0\rightarrow
1) {\partial\over\partial {r}_{1}^{\nu}}\right
\rbrack\omega({\bf
r},t-\tau)\frac{\omega}{\gamma}({\bf r}_1,t-\tau), \label{lb29b}
\end{eqnarray}
derived in \cite{pre,bbgky}, which is valid for flows that are not necessarily axisymmetric and not
necessarily Markovian \footnote{The Markovian approximation may not be
justified in every situation since it has been found numerically that
point vortices can exhibit long jumps (L\'evy flights) and strong
correlations \cite{kawahara,yoshida}. We also note that
Eq. (\ref{lb29b}) takes into account collective effects provided that
${\cal G}$ is the Green function associated with the {\it full}
equation (32) in Ref. \cite{bbgky} (this point was not mentioned in our
previous papers). If we neglect collective effects, ${\cal G}$ is
replaced by ${\cal G}_{bare}$ corresponding to the mere advection of
the vortices by the mean field. In that case, Eq. (\ref{lb29b}) directly leads to
Eq. (\ref{lb29}) \cite{bbgky,kindetail}.}. Equation (\ref{lb29}) is the counterpart of the
Landau equation in plasma physics.  The connection between Eqs. (\ref{lb27}) and (\ref{lb28}) is clear. If we neglect collective effects, Eq. (\ref{lb12}) reduces to ${\cal L}G_{bare}(n,r,r')=-\delta(r-r')/2\pi r$  (it does not depend on $\sigma$ anymore)  so that  $G_{bare}(n,r,r')$ is just the Fourier transform of the ``bare'' potential of interaction $u$ (i.e. the Green function of $L$). We thus have $G_{bare}(n,r,r')=\hat{u}_n(r,r')$ (see Appendix \ref{sec_pot}). When collective effects
are taken into account, the ``bare'' potential
$G_{bare}(n,r,r')=\hat{u}_n(r,r')$ is replaced by a ``dressed'' potential
$G(n,r,r',\sigma)$ without changing the overall structure of the
kinetic equation. This is similar to the case of plasma physics where
the bare potential $\hat{u}(k)^2$ in the Landau equation is replaced
by the dressed potential $\hat{u}(k)^2/|\epsilon({\bf k},{\bf k}\cdot
{\bf v})|^2$, including the dielectric function, in the Lenard-Balescu
equation. In plasma physics, collective effects are important because
they account for screening effects and regularize, at the scale of the
Debye length, the logarithmic divergence that occurs in the Landau
equation. In our context, there is no divergence in Eq. (\ref{lb29})
so that collective effects may be less crucial than in plasma
physics. So far, their influence is not clearly understood.

Finally, we note that the kinetic equation can be rewritten
\begin{equation}
\frac{\partial\omega}{\partial t}= 2\pi^2 \gamma \frac{1}{r}\frac{\partial}{\partial r}\int_{0}^{+\infty} r'\, dr'  \, \chi(r,r',t) \delta(\Omega(r,t)-\Omega(r',t))\left (\frac{1}{r}\frac{\partial}{\partial r}-\frac{1}{r'}\frac{\partial}{\partial r'}\right )\omega(r,t)\omega(r',t),
\label{lb30}
\end{equation}
with the notation
\begin{equation}
\chi(r,r',t)=\sum_{n} |n|  |G(n,r,r',n\Omega(r,t))|^2.
\label{lb31}
\end{equation}
When collective effects are neglected $\chi(r,r',t)$ is replaced by  $\chi_{bare}(r,r')=\sum_{n} |n| \hat{u}_n(r,r')^2$. For the usual potential of interaction (see Appendix \ref{sec_pot}), it reduces to  $\chi_{bare}(r,r')=-(1/8\pi^2)\ln\left\lbrack 1-(r_</r_>)^2\right\rbrack$.

\subsection{The relaxation time of the system as a whole}
\label{sec_relaxwhole}

The kinetic equation (\ref{lb30}) is valid at the order $1/N$ so that it describes the ``collisional'' evolution of the system on a timescale $\sim N t_D$. This kinetic equation conserves the total circulation $\Gamma$, the energy $E$ and the angular momentum $L$. It also monotonically increases the Boltzmann entropy $S$ ($H$-theorem). The Boltzmann distribution (\ref{pvm14}) is always a steady state of this kinetic equation, but it is not the only one: any vorticity distribution which is associated with a {\it monotonic} profile of angular velocity is a steady state of Eq. (\ref{lb30}). The kinetic equation admits therefore an infinite number of steady states! As explained in previous works, the collisional evolution of the point vortex gas (at the order $1/N$) is due to a condition of resonance between distant orbits of the point vortices. For axisymmetric systems, the condition of resonance, encapsulated in the $\delta$-function, corresponds to $\Omega(r',t)=\Omega(r,t)$ with $r'\neq r$ (the self-interaction at $r'=r$ does not produce transport since the term in parenthesis in Eq. (\ref{lb30}) vanishes identically). The vorticity profile $\omega(r,t)$  at position $r$ will change under the effects of ``collisions'' if there exists point vortices at $r'\neq r$ which rotate with the same angular velocity as point vortices located at $r$. This is possible only if the profile of angular velocity is non-monotonic \footnote{We recall that the system must be dynamically stable for the kinetic theory to be valid. It is experimentally observed that there exists vorticity profiles with a non-monotonic profile of angular velocity that are Euler stable \cite{dubin}.}. Therefore, the evolution is truly due to {\it distant} collisions between point vortices (this is strikingly different from the case of plasmas and stellar systems where the collisions are more ``local''). The evolution stops when the profile of angular velocity becomes monotonic (so that there is no resonance) even if the system has not reached the statistical equilibrium state given by the Boltzmann distribution. This ``kinetic blocking'' has been illustrated numerically in \cite{clemou}.  Distant collisions between point vortices have the tendency to create a monotonic profile of angular velocity. If the initial condition already has a monotonic profile of angular velocity, the collision term vanishes (at the order $1/N$) because there is no resonance. The kinetic equation reduces to
 \begin{equation}
\frac{\partial\omega}{\partial t}=0,
\label{rw1}
\end{equation}
so that the vorticity does not evolve at all on a timescale $\sim N t_D$. This implies that, for an axisymmetric distribution of point vortices, the relaxation time is larger than $Nt_D$. We therefore expect that
 \begin{equation}
t_{R}>Nt_D  \qquad ({\rm axisymmetric \,\, flows})
\label{rw2}
\end{equation}
Since the relaxation process is due to more complex correlations between point vortices, we have to develop the kinetic theory at higher orders (taking into account three-body, four-body,... correlation functions) in order to obtain the relaxation time. If the collision term does not vanish at the next order of the expansion, the kinetic theory would imply a relaxation time of the order $N^2t_D$ \cite{kindetail}. However, the problem could be more complicated and yield a larger relaxation time like $e^N t_D$. In fact, it is not even granted that the system will ever relax towards statistical equilibrium (the point vortex gas may be non-ergodic \cite{khanin}). By contrast, for non-axisymmetric flows, since there are potentially more resonances between point vortices [see the complicated kinetic equation (\ref{lb29b})], the relaxation time could be reduced and approach the natural scaling
\begin{equation}
t_{R}\sim Nt_D\qquad ({\rm non-axisymmetric\,\, flows})
\label{rw3}
\end{equation}
predicted by the first order kinetic theory. This linear scaling has
been observed for 2D point vortices with non-axisymmetric distribution
in \cite{kn}. However, very little is known concerning the properties
of Eq. (\ref{lb29b}) and its convergence (or not) towards the
Boltzmann distribution. It could approach the Boltzmann distribution
(since entropy increases), without reaching it exactly.

Similar results have been found for one-dimensional plasmas and
stellar systems and for the HMF model (see the conclusion for more
details). For spatially homogeneous one dimensional systems, the
Lenard-Balescu collision term vanishes (no resonance) so that the
relaxation time is larger than $Nt_D$. For spatially inhomogeneous one
dimensional systems, since there are potentially more resonances (this
can be seen by using angle-action variables \cite{angleaction}), the
relaxation time can be reduced and approach the natural scaling
$Nt_D$. For $d$-dimensional plasmas and stellar systems, with $d>1$,
there are always resonances since the condition is ${\bf k}\cdot {\bf
v}={\bf k}\cdot {\bf v}'$. The Lenard-Balescu collision term vanishes
only for the Boltzmann distribution and the relaxation time is
$t_{R}\sim Nt_D$ (or $t_R\sim (N/\ln N) t_D$ for stellar systems)
corresponding to the first order of the kinetic theory.

\section{Stochastic process of a test vortex: Diffusion and drift}
\label{sec_stoch}

\subsection{The Fokker-Planck equation}
\label{sec_fp}

In the previous section, we have studied the evolution of a system of point vortices as a whole. We now consider the relaxation of a test vortex is a ``sea'' of field vortices with a steady axisymmetric vorticity profile $\omega(r)$. The field vortices play the role of a bath. We assume that the field vortices are either at statistical equilibrium with the Boltzmann distribution (thermal bath), in which case their distribution does not change at all, or in a stable axisymmetric  steady state of the 2D Euler equation with a monotonic profile of angular velocity (as we have just seen, this profile does not change on a timescale of order $Nt_D$). We assume that the test vortex is initially located at a radial distance ${r}_0$ and we study how it progressively acquires the distribution of the field vortices due to distant ``collisions''. As we shall see, the test vortex has a stochastic motion and the evolution of the distribution function $P({r},t)$, the probability density of finding the test vortex at radial position ${r}$ at time $t$,  is governed by a Fokker-Planck equation involving a diffusion term and a drift term that can be analytically obtained. The Fokker-Planck equation may then be solved with the initial condition $P({r},0)=\delta({r}-{r}_0)/2\pi r$ to yield $P(r,t)$. This problem has been investigated in our previous papers \cite{preR,pre,bbgky,copenhaguen,kindetail}, but we shall give here a direct and more rigorous derivation of the coefficients or diffusion and drift, taking collective effects into account.

The equations of motion of the test vortex are
\begin{equation}
\label{fp1}
\frac{dr}{dt}=\frac{1}{r}\frac{\partial \delta\psi}{\partial\theta},\qquad \frac{d\theta}{dt}=\Omega(r)-\frac{1}{r}\frac{\partial \delta\psi}{\partial r}.
\end{equation}
They include the effect of the mean field $\Omega(r)$ which produces a zeroth-order net rotation plus a stochastic component $\delta\psi$ of order $1/\sqrt{N}$ (fluctuations) which takes into account the deviations from the mean field. They can be formally integrated as
\begin{equation}
\label{fp2}
r(t)=r+\int_{0}^{t}dt'\, \frac{1}{r(t')}\frac{\partial \delta\psi}{\partial\theta}(r(t'),\theta(t'),t'),
\end{equation}
\begin{equation}
\label{fp3}
\theta(t)=\theta+\int_{0}^{t}dt'\, \Omega(r(t'))-\int_{0}^{t}dt'\, \frac{1}{r(t')}\frac{\partial \delta\psi}{\partial r}(r(t'),\theta(t'),t'),
\end{equation}
where we have assumed that, initially, the test vortex is at $(r,\theta)$. Note that the ``initial'' time considered here does not necessarily coincide with the original time mentioned above. Since the fluctuation $\delta\psi$ of the stream function is a small quantity, the foregoing equations can be solved iteratively. At the order $1/N$, which corresponds to quadratic order in $\delta\psi$, we get for the radial distance
\begin{eqnarray}
\label{fp4}
r(t)=r+\int_{0}^{t}dt'\, \frac{1}{r}\frac{\partial \delta\psi}{\partial \theta}(r,\theta+\Omega t',t')+\int_{0}^{t}dt'\int_{0}^{t'}dt''\int_{0}^{t''}dt'''\,  \frac{1}{r^2}\frac{d\Omega}{dr}\frac{\partial^2 \delta\psi}{\partial \theta^2}(r,\theta+\Omega t',t')\frac{\partial \delta\psi}{\partial \theta}(r,\theta+\Omega t''',t''')\nonumber\\
-\int_{0}^{t}dt'\int_{0}^{t'}dt''\,  \frac{1}{r^2}\frac{\partial^2 \delta\psi}{\partial \theta^2}(r,\theta+\Omega t',t')\frac{\partial \delta\psi}{\partial r}(r,\theta+\Omega t'',t'')+\int_{0}^{t}dt'\int_{0}^{t'}dt''\,  \frac{1}{r^2}\frac{\partial^2 \delta\psi}{\partial r\partial \theta}(r,\theta+\Omega t',t')\frac{\partial \delta\psi}{\partial \theta}(r,\theta+\Omega t'',t'')\nonumber\\
-\int_{0}^{t}dt'\int_{0}^{t'}dt''\,  \frac{1}{r^3}\frac{\partial \delta\psi}{\partial \theta}(r,\theta+\Omega t',t')\frac{\partial \delta\psi}{\partial \theta}(r,\theta+\Omega t'',t'').\qquad
\end{eqnarray}
As the changes in the radial distance are small, the dynamics of the test vortex can be represented by a stochastic process governed by a Fokker-Planck equation \cite{risken}. If we denote by $P({r},t)$ the density probability of finding the test vortex at radial distance ${r}$ at time $t$, normalized such that $\int_0^{+\infty}P(r,t) 2\pi r\, dr=1$, the general form of this equation is
\begin{equation}
\label{fp5}
\frac{\partial P}{\partial t}=\frac{1}{r}\frac{\partial}{\partial r}\left ( r\frac{\partial}{\partial
r} DP\right )-\frac{1}{r}\frac{\partial}{\partial r}\left (rPA\right ).
\end{equation}
The diffusion coefficient and the drift term are given by
\begin{equation}
\label{fp6}
D(r)=\lim_{t\rightarrow +\infty}\frac{1}{2t} \langle (r(t)-r)^{2}\rangle,
\end{equation}
\begin{equation}
\label{fp7}
A(r)=\lim_{t\rightarrow +\infty}\frac{1}{t} \langle r(t)-r\rangle.
\end{equation}
In writing these limits, we have implicitly assumed that the time $t$ is long with respect to the fluctuation time but short with respect to the relaxation time (of order $Nt_D$), so that the expression (\ref{fp4}) can be used to evaluate Eqs. (\ref{fp6}) and (\ref{fp7}). As shown in our previous papers \cite{preR,pre,bbgky,copenhaguen,kindetail}, it is relevant to rewrite the Fokker-Planck equation in the alternative form
\begin{equation}
\label{fp8}
\frac{\partial P}{\partial t}=\frac{1}{r}\frac{\partial}{\partial r}\left \lbrack r \left (D\frac{\partial P}{\partial
r}- P V_{drift}\right )\right\rbrack.
\end{equation}
The total drift term is
\begin{equation}
\label{fp9}
A=V_{drift}+\frac{d D}{d r},
\end{equation}
where $V_{drift}$ is the ``essential'' part of the drift term, while the second term is due to the spatial variations of the diffusion coefficient. As we shall see, this decomposition arises naturally in the following analysis. The two expressions (\ref{fp5}) and (\ref{fp8}) have their own interest. The expression (\ref{fp5}) where the diffusion coefficient is placed after the second derivative $\partial^2(DP)$ involves the total drift $A$ and the expression (\ref{fp8}) where the diffusion coefficient is placed between the derivatives $\partial D\partial P$ isolates the essential part of the drift $V_{drift}$.  We shall see in Sec. \ref{sec_conn} that this second form is directly related to the kinetic equation (\ref{lb30}). It has therefore a clear physical interpretation.

We shall now calculate the diffusion coefficient and the drift term from Eqs. (\ref{fp6}) and (\ref{fp7}), using the results of Sec. \ref{sec_lb} that allow to take collective effects into account. Note that a similar calculation of these terms, directly from the equations of motion of a test vortex, has been made in Appendix C of \cite{clemou}, neglecting collective effects. The diffusion coefficient can also be obtained from the Kubo formula and the drift term from a linear response theory \cite{preR,pre,bbgky}.

\subsection{The diffusion coefficient}
\label{sec_diff}

We first compute the diffusion coefficient defined by Eq. (\ref{fp6}). Using Eq. (\ref{fp4}), we see that it is given, at the order $1/N$, by
\begin{equation}
D=\frac{1}{2r^2 t}\int_{0}^{t}dt'\int_{0}^{t}dt'' \, \left\langle \frac{\partial\delta\psi}{\partial\theta}(r,\theta+\Omega t',t')\frac{\partial\delta\psi}{\partial\theta}(r,\theta+\Omega t'',t'')\right\rangle.
\label{diff1}
\end{equation}
By the inverse Fourier-Laplace transform, we have
\begin{eqnarray}
\left\langle \frac{\partial\delta\psi}{\partial\theta}(r,\theta+\Omega t',t')\frac{\partial\delta\psi}{\partial\theta}(r,\theta+\Omega t'',t'')\right\rangle=-\sum_{n}\sum_{n'}\int_{\cal C}\frac{d\sigma}{2\pi}\int_{\cal C}\frac{d\sigma'}{2\pi} \, n n' e^{in(\theta+\Omega t')}e^{-i\sigma t'}e^{in'(\theta+\Omega t'')}e^{-i\sigma' t''} \nonumber\\
\times\langle \delta\tilde\psi(n,r,\sigma)\delta\tilde\psi(n',r,\sigma')\rangle.
\label{diff2}
\end{eqnarray}
Substituting Eq. (\ref{lb18}) in Eq. (\ref{diff2}), and carrying out the summation over $n'$ and the integrals over $\sigma'$ and $\sigma$, we end up with the result
\begin{eqnarray}
\left\langle \frac{\partial\delta\psi}{\partial\theta}(r,\theta+\Omega t',t')\frac{\partial\delta\psi}{\partial\theta}(r,\theta+\Omega t'',t'')\right\rangle=2\pi\gamma \sum_{n}\int_{0}^{+\infty}r'\, dr' \, n^2 e^{in(\Omega-\Omega')(t'-t'')}|G(n,r,r',n\Omega')|^2 \omega(r').
\label{diff3}
\end{eqnarray}
This expression shows that the correlation function appearing in Eq. (\ref{diff3}) only depends on $|t'-t''|$. Using the identity
\begin{eqnarray}
\int_{0}^{t}dt'\int_{0}^{t}dt''\, f(|t'-t''|)=2\int_{0}^{t}dt'\int_{0}^{t'}dt''\, f(|t'-t''|)=2\int_{0}^{t}ds\, (t-s)f(s),
\label{diff4}
\end{eqnarray}
we find, for $t\rightarrow +\infty$, that
\begin{equation}
D=\frac{1}{r^2}\int_{0}^{+\infty}\left\langle \frac{\partial\delta\psi}{\partial\theta}(r,\theta,0)\frac{\partial\delta\psi}{\partial\theta}(r,\theta+\Omega s,s)\right\rangle  \, ds.
\label{diff5}
\end{equation}
This is the Kubo formula for our problem. Replacing the correlation function by its expression (\ref{diff3}), we get
\begin{equation}
D=2\pi\gamma\frac{1}{r^2}\int_{0}^{+\infty}ds \sum_{n}\int_{0}^{+\infty}r'\, dr' \, n^2 e^{in(\Omega-\Omega')s}|G(n,r,r',n\Omega')|^2 \omega(r').
\label{diff6}
\end{equation}
Using the symmetry $s\rightarrow -s$, we can replace $\int_{0}^{+\infty}ds$ by $(1/2)\int_{-\infty}^{+\infty}ds$. Then, using the identity
\begin{equation}
\delta(\sigma)=\int_{-\infty}^{+\infty} e^{i \sigma t}\, \frac{dt}{2\pi},
\label{delta}
\end{equation}
and Eq. (\ref{lb22}), we obtain the final expression
\begin{equation}
D=2\pi^2\gamma\frac{1}{r^2}\sum_{n}\int_{0}^{+\infty}r'\, dr' \, |n|  |G(n,r,r',n\Omega)|^2 \delta(\Omega-\Omega') \omega(r').
\label{diff7}
\end{equation}

\subsection{The essential part of the drift term}
\label{sec_ess}

We now compute the drift term defined by Eq. (\ref{fp7}). We need to keep terms up to order $1/N$. From Eq. (\ref{fp4}), the first term to compute is
\begin{equation}
A_{I}=\frac{1}{r t}\int_{0}^{t}dt'\, \left\langle \frac{\partial\delta\psi}{\partial\theta}(r,\theta+\Omega t',t')\right\rangle.
\label{ess1}
\end{equation}
By the inverse Fourier-Laplace transform, we have
\begin{equation}
\left\langle \frac{\partial\delta\psi}{\partial\theta}(r,\theta+\Omega t',t')\right\rangle=i\sum_{n}\int_{\cal C}\frac{d\sigma}{2\pi} \, n  e^{i n(\theta+\Omega t')}e^{-i\sigma t'}\langle \delta\tilde\psi(n,r,\sigma)\rangle.
\label{ess2}
\end{equation}
Using Eq. (\ref{lb11}), we find that
\begin{equation}
\langle \delta\tilde\psi(n,r,\sigma)\rangle=\int_{0}^{+\infty}2\pi r'\, dr'\, \frac{\langle \delta\hat\omega(n,r',0)\rangle}{i(n\Omega'-\sigma)}G(n,r,r',\sigma).
\label{ess3}
\end{equation}
Now, using the fact that the test vortex is initially located in $(r,\theta)$, so that $\langle \delta\omega(\theta',r',0)\rangle=\gamma \delta(\theta'-\theta)\delta(r'-r)/r$,  we obtain from Eq. (\ref{lb8}), the result
\begin{equation}
\langle \delta\hat\omega(n,r',0)\rangle={\gamma}e^{-in\theta}\frac{\delta(r'-r)}{2\pi r}.
\label{ess4}
\end{equation}
Substituting these expressions in Eq. (\ref{ess2}), we get
\begin{equation}
\left\langle \frac{\partial\delta\psi}{\partial\theta}(r,\theta+\Omega t',t')\right\rangle={\gamma}\sum_{n}\int_{\cal C}\frac{d\sigma}{2\pi} \, n  e^{i (n \Omega-\sigma) t'}G(n,r,r,\sigma)\frac{1}{n\Omega-\sigma}.
\label{ess5}
\end{equation}
Therefore, the drift term (\ref{ess1}) is given by
\begin{equation}
A_{I}={\gamma}\frac{1}{r t}\int_{0}^{t}dt'\, \sum_{n}\int_{\cal C}\frac{d\sigma}{2\pi} \, n  e^{i (n \Omega-\sigma) t'}G(n,r,r,\sigma)\frac{1}{n\Omega-\sigma}.
\label{ess6}
\end{equation}
We now use the Plemelj formula to evaluate the integral over $\sigma$. The term corresponding to the imaginary part in the  Plemelj formula is
\begin{equation}
A_{I}^{(a)}=i\pi {\gamma}\frac{1}{r t}\int_{0}^{t}dt'\, \sum_{n}\int_{-\infty}^{+\infty}\frac{d\sigma}{2\pi} \, n  e^{i (n \Omega-\sigma) t'}G(n,r,r,\sigma)\delta(n\Omega-\sigma).
\label{ess7}
\end{equation}
Integrating over $\sigma$ and $t'$, we obtain
\begin{equation}
A_{I}^{(a)}=-\frac{\gamma}{2r} \sum_{n} \, n  \, {\rm Im}\, G(n,r,r,n\Omega).
\label{ess8}
\end{equation}
The term corresponding to the real part in the Plemelj formula  is
\begin{equation}
A_{I}^{(b)}= {\gamma}\frac{1}{r t}\int_{0}^{t}dt'\, \sum_{n}{\cal P}\int_{-\infty}^{+\infty}\frac{d\sigma}{2\pi} \, n  e^{i (n \Omega-\sigma) t'}G(n,r,r,\sigma)\frac{1}{n\Omega-\sigma}.
\label{ess10}
\end{equation}
Integrating over $t'$, we can convert this expression to the form
\begin{equation}
A_{I}^{(b)}= -i\frac{\gamma}{r} \sum_{n}{\cal P}\int_{-\infty}^{+\infty}\frac{d\sigma}{2\pi} \, n  G(n,r,r,\sigma)\frac{1}{(n\Omega-\sigma)^2}\frac{1}{t}\left\lbrace i\sin \left\lbrack (n\Omega-\sigma)t\right\rbrack+\cos \left\lbrack (n\Omega-\sigma)t\right\rbrack-1\right\rbrace.
\label{ess11}
\end{equation}
For $t\rightarrow +\infty$, using the identity (as in Appendix C of \cite{clemou}):
\begin{equation}
\lim_{t\rightarrow +\infty}\frac{1-\cos(tx)}{t x^2}=\pi\delta(x),
\label{ess12}
\end{equation}
and integrating over $\sigma$, we obtain
\begin{equation}
A_{I}^{(b)}= -\frac{\gamma}{2 r} \sum_{n} \, n \, {\rm Im}\, G(n,r,r,n\Omega),
\label{ess13}
\end{equation}
which is the same as Eq. (\ref{ess8}). Finally, summing Eqs. (\ref{ess8}) and (\ref{ess13}) and using Eq. (\ref{lb25}), we find that
\begin{equation}
A_{I}=2\gamma \pi^2\frac{1}{r} \sum_{n} \int_{0}^{+\infty} dr' \, |n|  |G(n,r,r',n\Omega)|^2 \delta(\Omega-\Omega')\frac{d\omega'}{d r'}.
\label{ess14}
\end{equation}
As we shall see, it corresponds to the ``essential'' part of the drift term denoted $V_{drift}$ in Eq. (\ref{fp9}).

\subsection{The part of the drift term due to the spatial inhomogeneity of the diffusion coefficient}
\label{sec_add}

In the evaluation of the total drift, at the order $1/N$, the second term to compute is
\begin{equation}
A_{II}=\frac{1}{t}\int_{0}^{t}dt'\int_{0}^{t'}dt''\int_{0}^{t''}dt'''\, \frac{1}{r^2}\frac{d\Omega}{dr}\left\langle \frac{\partial^2\delta\psi}{\partial\theta^2}(r,\theta+\Omega t',t') \frac{\partial\delta\psi}{\partial\theta}(r,\theta+\Omega t''',t''')  \right\rangle.
\label{add1}
\end{equation}
By the inverse Fourier-Laplace transform, we have
\begin{eqnarray}
\left\langle \frac{\partial^2\delta\psi}{\partial\theta^2}(r,\theta+\Omega t',t')\frac{\partial\delta\psi}{\partial\theta}(r,\theta+\Omega t''',t''')\right\rangle=-i\sum_{n}\sum_{n'}\int_{\cal C}\frac{d\sigma}{2\pi}\int_{\cal C}\frac{d\sigma'}{2\pi} \, n^2 n' e^{in(\theta+\Omega t')}e^{-i\sigma t'}e^{in'(\theta+\Omega t''')}e^{-i\sigma' t'''} \nonumber\\
\times\langle \delta\tilde\psi(n,r,\sigma)\delta\tilde\psi(n',r,\sigma')\rangle.
\label{add2}
\end{eqnarray}
Substituting Eq. (\ref{lb18}) in Eq. (\ref{add2}), and carrying out the summation over $n'$ and the integrals over $\sigma'$ and $\sigma$, we end up with the result
\begin{eqnarray}
\left\langle \frac{\partial^2\delta\psi}{\partial\theta^2}(r,\theta+\Omega t',t')\frac{\partial\delta\psi}{\partial\theta}(r,\theta+\Omega t''',t''')\right\rangle=i\, 2\pi \gamma \sum_{n}\int_{0}^{+\infty}r'\, dr' \, n^3 e^{in(\Omega-\Omega')(t'-t''')}|G(n,r,r',n\Omega')|^2 \omega(r').
\label{add3}
\end{eqnarray}
This expression shows that the correlation function appearing in Eq. (\ref{add1}) only depends on $|t'-t'''|$. Using the identity
\begin{eqnarray}
\int_{0}^{t'}dt''\int_{0}^{t''}dt'''\, f(|t'-t'''|)=\int_{0}^{t'}dt''' \, (t'-t''')f(|t'-t'''|),
\label{add4}
\end{eqnarray}
we find that
\begin{equation}
A_{II}=i \, 2\pi \gamma\frac{1}{t}\int_{0}^{t}dt'\int_{0}^{t'}dt'''\sum_{n}\int_{0}^{+\infty}r'\, dr' \, (t'-t''')n^3 \frac{1}{r^2}\frac{d\Omega}{dr} e^{in(\Omega-\Omega')(t'-t''')}|G(n,r,r',n\Omega')|^2 \omega(r').
\label{add5}
\end{equation}
This can be rewritten
\begin{equation}
A_{II}=2\pi\gamma\frac{1}{t}\int_{0}^{t}dt'\int_{0}^{t'}dt'''\sum_{n}\int_{0}^{+\infty}r'\, dr' \, n^2 \frac{1}{r^2} \frac{\partial}{\partial r}\left (e^{in(\Omega-\Omega')(t'-t''')}\right ) |G(n,r,r',n\Omega')|^2 \omega(r'),
\label{add6}
\end{equation}
or, equivalently,
\begin{eqnarray}
A_{II}=2\pi\gamma\frac{1}{t}\frac{\partial}{\partial r}\int_{0}^{t}dt'\int_{0}^{t'}dt'''\sum_{n}\int_{0}^{+\infty}r'\, dr' \, n^2 \frac{1}{r^2}e^{in(\Omega-\Omega')(t'-t''')} |G(n,r,r',n\Omega')|^2 \omega(r')\nonumber\\
-2\pi\gamma\frac{1}{t}\int_{0}^{t}dt'\int_{0}^{t'}dt'''\sum_{n}\int_{0}^{+\infty}r'\, dr' \, n^2 e^{in(\Omega-\Omega')(t'-t''')}\frac{\partial}{\partial r} \left (\frac{1}{r^2} |G(n,r,r',n\Omega')|^2\right ) \omega(r').
\label{add7}
\end{eqnarray}
Since the integrand only depends on $|t'-t'''|$, using the identity (\ref{diff4}), we obtain for $t\rightarrow +\infty$,
\begin{eqnarray}
A_{II}=2\pi\gamma\frac{\partial}{\partial r}\int_{0}^{+\infty}ds\sum_{n}\int_{0}^{+\infty}r'\, dr' \, n^2 \frac{1}{r^2}e^{in(\Omega-\Omega')s} |G(n,r,r',n\Omega')|^2 \omega(r')\nonumber\\
-2\pi\gamma\int_{0}^{+\infty}ds\sum_{n}\int_{0}^{+\infty}r'\, dr' \, n^2 e^{in(\Omega-\Omega')s}\frac{\partial}{\partial r} \left (\frac{1}{r^2} |G(n,r,r',n\Omega')|^2\right ) \omega(r').
\label{add8}
\end{eqnarray}
Using the symmetry $s\rightarrow -s$, we can replace $\int_{0}^{+\infty}ds$ by $(1/2)\int_{-\infty}^{+\infty}ds$. Then, using the identities (\ref{delta}) and (\ref{lb22}), we obtain the expression
\begin{eqnarray}
A_{II}=2\pi^2\gamma\frac{\partial}{\partial r}\sum_{n}\int_{0}^{+\infty}r'\, dr' \, |n| \frac{1}{r^2}\delta(\Omega-\Omega') |G(n,r,r',n\Omega')|^2 \omega(r')\nonumber\\
-2\pi^2\gamma\sum_{n}\int_{0}^{+\infty}r'\, dr' \, |n| \delta(\Omega-\Omega')\frac{\partial}{\partial r} \left (\frac{1}{r^2} |G(n,r,r',n\Omega')|^2\right ) \omega(r').
\label{add9}
\end{eqnarray}
In the first term, we recover the diffusion coefficient (\ref{diff7}) so that finally
\begin{equation}
A_{II}=\frac{dD}{dr}-2\pi^2\gamma\sum_{n}\int_{0}^{+\infty}r'\, dr' \, |n| \delta(\Omega-\Omega')\frac{\partial}{\partial r} \left (\frac{1}{r^2} |G(n,r,r',n\Omega')|^2\right ) \omega(r').
\label{add10}
\end{equation}

The third term to compute is
\begin{equation}
A_{III}=-\frac{1}{r^3t}\int_{0}^{t}dt'\int_{0}^{t'}dt''\, \left\langle \frac{\partial\delta\psi}{\partial\theta}(r,\theta+\Omega t',t') \frac{\partial\delta\psi}{\partial\theta}(r,\theta+\Omega t'',t'')  \right\rangle.
\label{add11}
\end{equation}
This term is just proportional to the diffusion coefficient (\ref{diff1}) so we get
\begin{equation}
A_{III}=-\frac{2}{r}D=-4\pi^2\gamma\frac{1}{r^3}\sum_{n}\int_{0}^{+\infty}r'\, dr' \, |n|  |G(n,r,r',n\Omega)|^2 \delta(\Omega-\Omega') \omega(r').
\label{add12}
\end{equation}

Finally, the fourth and fifth terms to compute are
\begin{equation}
A_{IV}=-\frac{1}{r^2 t}\int_{0}^{t}dt'\int_{0}^{t'}dt''\, \left\langle \frac{\partial^2\delta\psi}{\partial\theta^2}(r,\theta+\Omega t',t') \frac{\partial\delta\psi}{\partial r}(r,\theta+\Omega t'',t'')  \right\rangle,
\label{add13}
\end{equation}
and
\begin{equation}
A_{V}=\frac{1}{r^2t}\int_{0}^{t}dt'\int_{0}^{t'}dt''\, \left\langle \frac{\partial^2\delta\psi}{\partial r\partial \theta}(r,\theta+\Omega t',t') \frac{\partial\delta\psi}{\partial\theta}(r,\theta+\Omega t'',t'')  \right\rangle.
\label{add14}
\end{equation}
Substituting the inverse Fourier-Laplace transform of the fluctuations of the stream function in these equations, and summing the resulting expressions using the fact that the correlation function of the fluctuations of the stream function is proportional to $\delta_{n,-n'}$, we obtain
\begin{equation}
A_{IV}+A_{V}=\frac{1}{r^2 t}\int_{0}^{t}dt'\int_{0}^{t'}dt''\sum_{n}\sum_{n'}\int_{\cal C}\frac{d\sigma}{2\pi}\int_{\cal C}\frac{d\sigma'}{2\pi} \, n^2 e^{in(\theta+\Omega t')}e^{-i\sigma t'}e^{in'(\theta+\Omega t'')}e^{-i\sigma' t''} \frac{\partial}{\partial r}\langle \delta\tilde\psi(n,r,\sigma)\delta\tilde\psi(n',r,\sigma')\rangle.
\label{add15}
\end{equation}
Substituting Eq. (\ref{lb18}) in Eq. (\ref{add15}), and carrying out the summation over $n'$ and the integrals over $\sigma'$ and $\sigma$, we end up with the result
\begin{eqnarray}
A_{IV}+A_{V}=2\pi\gamma\frac{1}{r^2 t}\int_{0}^{t}dt'\int_{0}^{t'}dt''\sum_{n}\int_{0}^{+\infty}r'\, dr' \, \frac{n^2}{r^2} e^{in(\Omega-\Omega')(t'-t'')} \frac{\partial}{\partial r} (|G(n,r,r',n\Omega')|^2) \omega(r').
\label{add16}
\end{eqnarray}
This expression shows that the correlation function appearing under the integral sign only depends on $|t'-t''|$. Using the identity (\ref{diff4}) we find, for $t\rightarrow +\infty$, that
\begin{eqnarray}
A_{IV}+A_{V}=2\pi\gamma\frac{1}{r^2}\int_{0}^{+\infty}ds\sum_{n}\int_{0}^{+\infty}r'\, dr' \, n^2 e^{in(\Omega-\Omega')s} \frac{\partial}{\partial r} (|G(n,r,r',n\Omega')|^2) \omega(r').
\label{add17}
\end{eqnarray}
Using the symmetry $s\rightarrow -s$, we can replace $\int_{0}^{+\infty}ds$ by $(1/2)\int_{-\infty}^{+\infty}ds$. Then, using the identities (\ref{delta}) and (\ref{lb22}), we obtain the expression
\begin{equation}
A_{IV}+A_{V}=2\pi^2\gamma\frac{1}{r^2}\sum_{n}\int_{0}^{+\infty}r'\, dr' \, |n|  \frac{\partial}{\partial r} (|G(n,r,r',n\Omega)|^2) \delta(\Omega-\Omega') \omega(r').
\label{add18}
\end{equation}
Finally, summing Eqs. (\ref{add10}), (\ref{add12}) and (\ref{add18}), we get
\begin{equation}
A_{II}+A_{III}+A_{IV}+A_{V}=\frac{dD}{dr}.
\label{add19}
\end{equation}
Therefore, recalling Eq. (\ref{ess14}), the complete expression of the drift term is
\begin{equation}
A=2\pi^2\gamma \frac{1}{r} \sum_{n} \int_{0}^{+\infty} dr' \, |n|  |G(n,r,r',n\Omega)|^2 \delta(\Omega-\Omega')\frac{d\omega'}{d r'}+\frac{dD}{d r}.
\label{add20}
\end{equation}

\subsection{Connection between the kinetic equation (\ref{lb30}) and the Fokker-Planck equation (\ref{fp8})}
\label{sec_conn}

We have established that the diffusion coefficient and the drift term are given by Eqs. (\ref{diff7}) and (\ref{add20}). Introducing the notation (\ref{lb31}), they can be written
\begin{equation}
D={2\pi^2\gamma}\frac{1}{r^2}\int_{0}^{+\infty}r'\, dr' \, \chi(r,r') \delta(\Omega-\Omega') \omega(r'),
\label{conn1}
\end{equation}
and
\begin{equation}
A={2\pi^2\gamma}\frac{1}{r} \int_{0}^{+\infty} dr' \, \chi(r,r') \delta(\Omega-\Omega')\frac{d \omega}{d r}(r')+\frac{dD}{dr}.
\label{conn2}
\end{equation}
Comparing Eq. (\ref{conn2}) with Eq. (\ref{fp9}), we see that the ``essential'' part of the drift term is
\begin{equation}
V_{drift}={2\pi^2\gamma}\frac{1}{r} \int_{0}^{+\infty} dr' \, \chi(r,r') \delta(\Omega-\Omega')\frac{d \omega}{d r}(r').
\label{conn3}
\end{equation}
On the other hand, using an integration by part in the first term of Eq. (\ref{conn2}), the total drift can be written
\begin{equation}
A=2\pi^2 \gamma \int_{0}^{+\infty} r r' dr' \, \omega(r')\left (\frac{1}{r}\frac{\partial}{\partial r}-\frac{1}{r'}\frac{\partial}{\partial r'}\right )\chi(r,r') \delta(\Omega-\Omega')\frac{1}{r^2}.
\label{conn4}
\end{equation}
Finally, using Eqs. (\ref{conn1}) and (\ref{conn3}), we find that the Fokker-Planck equation (\ref{fp8}) becomes
\begin{equation}
\frac{\partial P}{\partial t}= 2\pi^2 \gamma \frac{1}{r}\frac{\partial}{\partial r}\int_{0}^{+\infty} r'\, dr'  \, \chi(r,r') \delta(\Omega-\Omega')\left (\frac{1}{r}\omega'\frac{\partial P}{\partial r}-\frac{1}{r'}P\frac{d\omega'}{d r'}\right ).
\label{conn5}
\end{equation}
When collective effects are neglected, we recover the results obtained
in \cite{preR,pre,bbgky,copenhaguen,kindetail} by a different
method. As observed in our previous works, we note that the form of
Eq. (\ref{conn5}) is very similar to the form of
Eq. (\ref{lb30}). This shows that the Fokker-Planck equation
(\ref{conn5}), with the diffusion coefficient (\ref{conn1}) and the
drift term (\ref{conn3}), can be directly obtained from the kinetic
equation (\ref{lb30}) by replacing the time dependent distribution
$\omega(r',t)$ by the {\it static} distribution $\omega(r')$ of the
bath. This procedure transforms an integro-differential equation
(\ref{lb30}) into a differential equation (\ref{conn5}) \cite{bbgky}.
Although natural, the rigorous justification of this procedure
requires the detailed calculation that we have given here. In fact, we
can understand this result in the following manner. Equations
(\ref{lb30}) and (\ref{conn5}) describe the evolution of the
distribution function of a test vortex (described by the coordinate
$r$) interacting with field vortices (described by the running
coordinate $r'$). In Eq. (\ref{lb30}), all the vortices are
equivalent so that the distribution of the field vortices
$\omega(r',t)$ changes with time exactly like the distribution of the
test vortex $\omega(r,t)$. In Eq. (\ref{conn5}), the test vortex
and the field vortices are not equivalent since the field vortices
form a ``bath''. The field vortices have a steady (given)
distribution $\omega(r')$ while the distribution of the test vortex
$\omega(r,t)=N\gamma P(r,t)$ changes with time. This distinction was
particularly visible in our first derivation of the kinetic and
Fokker-Planck equations using projection operator methods \cite{pre}.

\subsection{Monotonic profile of angular velocity}
\label{sec_m}

If the profile of angular velocity $\Omega(r)$ of the field vortices
is monotonic, then using the identity
$\delta(\Omega(r)-\Omega(r'))=\delta(r-r')/|\Omega'(r)|$, we find that the expressions of the diffusion coefficient and of the drift  simplify into
\begin{eqnarray}
D(r)={2\pi^2\gamma}
\frac{\chi(r,r)}{|\Sigma(r)|}\omega(r),\label{m1}
\end{eqnarray}
and
\begin{eqnarray}
V_{drift}={2\pi^2\gamma}
\frac{\chi(r,r)}{|\Sigma(r)|}\frac{d\omega}{dr}(r),\label{m2}
\end{eqnarray}
where $\Sigma(r)=r\Omega'(r)$ is the local shear created by the field vortices. Comparing Eqs. (\ref{m1}) and (\ref{m2}), we obtain
\begin{eqnarray}
V_{drift}=D(r)\frac{d\ln\omega}{dr}.\label{m3}
\end{eqnarray}
This relation is valid for an arbitrary distribution of the field vortices provided that the profile of angular momentum is monotonic so that their distribution is steady (this important relation was first given in Eq. (123) of Ref. \cite{pre}). It can be viewed as a generalization of  the Einstein relation (see next section) for a bath that is out-of-equilibrium.  Therefore, the Fokker-Planck equation (\ref{fp8}) can be written
\begin{equation}
\label{m4}{\partial P\over\partial
t}=\frac{1}{r}{\partial\over\partial r}\biggl\lbrack r D(r)\biggl
({\partial P\over\partial r}-P\frac{d\ln\omega}{dr}\biggr
)\biggr\rbrack.
\end{equation}
This is a drift-diffusion equation describing the evolution of the distribution $P(r,t)$ of the test vortex in an ``effective potential'' $U_{eff}=-\ln\omega(r)$ produced by the field vortices. From this equation, we find that, for $t\rightarrow +\infty$, the distribution of the test vortex becomes equal to the distribution of the bath: $P_e(r)=\omega(r)/N\gamma$ at equilibrium.

If we neglect collective effects, we recover the results obtained in \cite{preR,pre,bbgky,copenhaguen,kindetail}. For the usual potential of interaction, the function $\chi_{bare}(r,r)$ is given by (see Appendix \ref{sec_pot}):
\begin{equation}
\label{m5}\chi_{bare}(r,r)=\frac{1}{8\pi^2}\ln\Lambda,
\end{equation}
where $\ln\Lambda=\sum_{m=1}^{+\infty}{1}/{m}$ is a logarithmically diverging Coulomb factor that has
to be regularized appropriately (see below).
Combining Eqs. (\ref{m1}), (\ref{m2}) and (\ref{m5}), the diffusion coefficient and the drift term can be rewritten
\begin{eqnarray}
D(r)=\frac{\gamma}{4}
\frac{\ln \Lambda}{|\Sigma(r)|}\omega(r),\qquad V_{drift}=\frac{\gamma}{4}
\frac{\ln\Lambda}{|\Sigma(r)|}\frac{d\omega}{dr}(r).\label{m7}
\end{eqnarray}
These expressions were first derived for a thermal bath in \cite{preR}
and extended to an arbitrary (steady Euler stable) distribution of the
field vortices in \cite{pre}. They also appear in the works
\cite{dubinjin,dubin2,sdprl,sd2} with a different interpretation (see
the conclusion for more details). An important feature of these
results is that the diffusion coefficient and the drift term {\it are
inversely proportional to the local shear}, a feature first noted in
\cite{preR}. Furthermore, the diffusion coefficient is proportional to
the vorticity and the drift term is proportional to the gradient of
the vorticity. Since ${\bf V}_{drift}=(D/\omega)\nabla\omega$ in
vectorial form, we find that, due to the drift, the test vortex
ascends the vorticity gradient. This is a purely deterministic effect
that appears when the distribution of the field vortices is spatially
inhomogeneous (in the absence of vorticity gradient, there is no
drift). As shown in \cite{preR,pre} (see also \cite{sdprl,sd2}), the
systematic drift of the test vortex is due to a polarization process:
the test vortex modifies the distribution of the field vortices and,
in response, the retroaction of this perturbation causes the drift of
the test vortex. In the absence of fluctuation, the test vortex would
reach a maximum of background vorticity where $\nabla\omega={\bf
0}$. In fact, this systematic effect is counterbalanced by the
diffusion term that tends to disperse the test vortex. Finally, an
equilibrium state results from these two antagonistic effects in which
the distribution of the test vortex coincides with the distribution of
the bath. This equilibration process is governed by
the Fokker-Planck equation (\ref{m4}).

The logarithmic divergence of the Coulomb factor $\ln\Lambda$ (that
persists if collective effects are included) was first noted in
\cite{preR}. It is due to the failure of the kinetic theory at small
scales where collisions between vortices are
strong. Phenomenologically, the logarithmic divergence can be
regularized by introducing cut-offs so that $\ln\Lambda=\ln(r/d)$
\footnote{A similar logarithmic divergence at small scales arises in
3D Coulombian plasmas and stellar systems. In that case, the Coulombian factor
has to be regularized at the Landau length.}.  A detailed calculation
of the lower cut-off $d$ has been made by Dubin \& Jin \cite{dubinjin,dubin2}. They
propose to take $d={\rm Max}(\delta,l)$ where $l$ is the trapping
distance $l=(\gamma/4\pi|\Sigma|)^{1/2}$ and $\delta$ is the
diffusion-limited maximum separation $\delta=(4D/|\Sigma|)^{1/2}$
where $D$ is the diffusion coefficient given by Eq. (\ref{m7}). Orders
of magnitude indicate that $r/l\sim R(|\Sigma|/\gamma)^{1/2}\sim
R(\Gamma/\gamma R^2)^{1/2}\sim N^{1/2}$ (where $R$ is the
system size), $\delta/l\sim (D/\gamma)^{1/2}\sim (\ln
\Lambda)^{1/2}\sim [\ln(\ln N)]^{1/2}$ and $r/\delta\sim
N^{1/2}/[\ln(\ln N)]^{1/2}$. Therefore, at leading order, the
Coulombian factor scales with $N$ like \cite{clemou}:
\begin{equation}
\label{m6}\ln\Lambda\sim \frac{1}{2}\ln N.
\end{equation}
This expression may be substituted in Eq. (\ref{m7}) if we are interested in orders of magnitude.

\subsection{Thermal bath: Boltzmann distribution}
\label{sec_tb}

For a thermal bath, the field vortices have the Boltzmann distribution of statistical equilibrium
\begin{eqnarray}
\omega({r})=A e^{-\beta \gamma \psi_{*}(r)}, \label{tb1}
\end{eqnarray}
where $\psi_{*}(r)=\psi(r)+({\Omega_{L}}/{2})r^{2}$ is the relative stream function taking into account the invariance by rotation of the system. We note the identity
\begin{eqnarray}
\frac{d\omega}{dr}=-\beta\gamma
\omega\frac{d\psi_{*}}{dr}=\beta\gamma\omega (\Omega-\Omega_{L})r,\label{tb2}
\end{eqnarray}
where we have used $-d\psi/dr=\Omega r$. Substituting this relation in Eq. (\ref{conn3}), we obtain
\begin{eqnarray}
V_{drift}=2\pi^{2}\gamma^2\beta\frac{1}{r}\int_{0}^{+\infty}dr'
\chi(r,r')\delta(\Omega-\Omega')\omega'(\Omega'-\Omega_L)r'.\label{tb3}
\end{eqnarray}
Using the $\delta$-function to replace $\Omega'$ by $\Omega$, then using
$\Omega-\Omega_{L}=-(1/r){d\psi_{*}}/{dr}$,  and comparing the resulting
expression with Eq. (\ref{conn1}), we finally find that
\begin{eqnarray}
V_{drift}=-D\beta\gamma \frac{d\psi_{*}}{dr}. \label{tb4}
\end{eqnarray}
In vectorial form, this can be written ${\bf V}_{drift}=-D\beta \gamma\nabla\psi_{*}$ \cite{preR}.
The drift is perpendicular to the relative mean field velocity
${\bf u}_{*}=-{\bf z}\times\nabla\psi_{*}$ and the
drift coefficient (mobility) satisfies an Einstein relation
\begin{eqnarray}
\xi=D\beta\gamma. \label{tb5}
\end{eqnarray}
We note that the drift coefficient $\xi$ and the diffusion coefficient $D$ depend on the position $r$ of the test vortex and we recall that the inverse temperature $\beta$ is negative in cases of physical interest. We stress that the Einstein relation is valid for the ``essential''
drift $V_{drift}$, not for the total drift $A$ that has a more complex expression due to the term $dD/dr$.  We do not have this subtlety for the usual Brownian motion where the diffusion coefficient is constant. For a thermal bath, using
Eq. (\ref{tb4}), the Fokker-Planck equation (\ref{fp8}) takes the
form
\begin{equation}
\label{tb6}{\partial P\over\partial
t}=\frac{1}{r}{\partial\over\partial r}\biggl\lbrack r D(r)\biggl
({\partial P\over\partial r}+\beta\gamma P\frac{d\psi_{*}}{dr}\biggr
)\biggr\rbrack,
\end{equation}
where $D(r)$ is given by Eq. (\ref{conn1}) with Eq. (\ref{tb1}). This
equation has a form similar to the familiar Smoluchowski equation in
Brownian theory. This is a drift-diffusion equation describing the
evolution of the distribution $P(r,t)$ of the test vortex in an
``effective potential'' $U_{eff}=\psi_{*}$ produced by the field
vortices. For $t\rightarrow +\infty$, the distribution of the test
vortex relaxes towards the Boltzmann distribution:
$P_e(r)=(A/N\gamma)e^{-\beta \gamma \psi_{*}(r)}$ at equilibrium. Of
course, if the profile of angular velocity of the Boltzmann
distribution is monotonic, we find that Eq. (\ref{m2}) with
Eq. (\ref{tb1}) returns Eq. (\ref{tb4}) with a diffusion coefficient
given by Eq. (\ref{m1}) with Eq. (\ref{tb1}). However,
Eqs. (\ref{tb4}) and (\ref{conn1}) are valid even if the profile of
angular velocity in the Boltzmann distribution is non-monotonic.
Finally, we note that the systematic drift $ {\bf
V}_{drift}=-D(r)\beta\gamma\nabla\psi_{*}$ of a point vortex
\cite{preR} is the counterpart of the dynamical friction ${\bf
F}_{fric}=-D_{\|}(v)\beta m {\bf v}$ of a star in a cluster
\cite{chandra1}, and the Smoluchowski-type Fokker-Planck equation (\ref{tb6}) is the
counterpart of the Kramers-Chandrasekhar equation (see \cite{kindetail} for a development of this
analogy).

\subsection{Gaussian vortex}
\label{sec_gaussian}

For illustration, we consider the particular case of a Gaussian distribution of field vortices
\begin{eqnarray}
\omega({r})=\frac{\Gamma\gamma\lambda}{2\pi} e^{-\lambda \gamma r^2/2}. \label{g1}
\end{eqnarray}
It can be viewed as a particular statistical equilibrium state of the form (\ref{tb1}) corresponding to $\beta\rightarrow 0$ and $\Omega_L\rightarrow +\infty$ in such a way that the product $\lambda\equiv \beta\Omega_L$ remains finite \cite{lp}. It is also similar to the familiar Maxwellian distribution of the velocities in a gas at inverse temperature $\lambda$ (provided that $r$ is replaced by $v$ and $\gamma$ by $m$). The Fokker-Planck equation (\ref{m4}) or (\ref{tb6}) describing the relaxation of a test vortex in a Gaussian bath is
\begin{equation}
\label{g2}{\partial P\over\partial
t}=\frac{1}{r}{\partial\over\partial r}\biggl\lbrack r D(r)\biggl
({\partial P\over\partial r}+\lambda\gamma Pr\biggr
)\biggr\rbrack,
\end{equation}
where $D(r)$ is given by Eq. (\ref{m1}) with Eq. (\ref{g1}). This
equation has a form similar to the familiar Kramers equation in
Brownian theory (provided that $r$ is replaced by $v$ and $\gamma$ by
$m$). This is a
drift-diffusion equation describing the evolution of the distribution
$P(r,t)$ of the test vortex in an ``effective potential''
$U_{eff}=r^2/2$ (quadratic) produced by the field vortices. For
$t\rightarrow +\infty$, the distribution of the test vortex relaxes
towards the Maxwellian distribution of the bath:
$P_e(r)=(\lambda\gamma/2\pi) e^{-\lambda \gamma r^2/2}$ at
equilibrium.

Using $\Omega(r)=r^{-2}\int_0^r \omega(r')r'\, dr'$ (see Sec. 6.1 of \cite{clemou}), the profile of angular velocity of the bath is
\begin{equation}
\label{g3}
\Omega(r)=\frac{\Gamma}{2\pi r^2}(1-e^{-\lambda\gamma r^2/2}).
\end{equation}
If we neglect collective effects, the diffusion coefficient is given by Eq. (\ref{m7}). Using Eqs. (\ref{g1}) and (\ref{g3}), we obtain the expression
\begin{equation}
\label{g4}
D(r)=\frac{\gamma^2\lambda\ln\Lambda r^2}{4(2e^{\lambda\gamma r^2/2}-\lambda\gamma r^2-2)}.
\end{equation}
We first note that the diffusion coefficient diverges like $D(r)\sim \ln\Lambda/(\lambda r^2)$ for $r\rightarrow 0$ due to the vanishing of the shear in the core of the vortex: $\Sigma(r)\sim -(\lambda^2\gamma^2\Gamma/8\pi)r^2$ as $r\rightarrow 0$. This indicates a failure of the kinetic theory for $r\rightarrow 0$. In fact, the kinetic theory developed in this paper is valid for sufficiently large shears. In the absence of shear, the expression of the diffusion coefficient is different (and finite) as discussed in \cite{chavsire1,houchesPH}. Therefore, the expression (\ref{g4}) of the diffusion coefficient is valid for sufficiently large $r$.  We note that the diffusion coefficient decays very rapidly at large distances since $D(r)\sim (\gamma^2\lambda\ln\Lambda/8) r^2e^{-\lambda\gamma r^2/2}$ for $r\rightarrow +\infty$. More generally, using Eq. (\ref{m1}) and the results of Sec. 6.1. of \cite{clemou}, we have $D(r)\sim (2\pi^3\gamma/\Gamma)\chi(r,r)\omega(r)r^2$ for $r\rightarrow +\infty$ for an arbitrary distribution $\omega(r)$ of the field vortices.

The Fokker-Planck equation (\ref{g2}) with the diffusion coefficient (\ref{g4}) has been studied in \cite{clemou} by applying results previously obtained in the context of the HMF model. For this model, the equivalent of the Fokker-Planck equation (\ref{g2}) with Eq. (\ref{g4}), where the position $r$ is replaced by the velocity $v$, has been studied in \cite{bd,clpre}. This Fokker-Planck equation presents unusual features because the diffusion coefficient decreases very rapidly with the distance. This leads to  ``anomalies'' with respect to the usual Brownian motion. By applying the approach of Bouchet \& Dauxois \cite{bd}, one finds \cite{clemou} that the auto-correlation function of the test vortex   $\langle r(0)r(r)\rangle$ decays algebraically like $\ln t/t$ (this algebraic decay was first obtained by Marksteiner {\it et al.} \cite{marksteiner} for the logarithmic Fokker-Planck equation to which Eq. (\ref{g2}) can be mapped).  On the other hand, by applying the approach of Chavanis \& Lemou \cite{clpre}, one finds \cite{clemou} that the normalized distribution $u(r,t)=P(r,t)/P_e(r)$ has a front structure and that the front evolves very slowly with time, scaling like $r_f(t)\propto (\ln t)^{1/2}$ for $t\rightarrow +\infty$. These results  show that the relaxation of $P(r,t)$ towards the equilibrium distribution $P_e(r)$ is not exponential. This is intrinsically  due to the absence of gap \cite{marksteiner,bd,clemou} in the spectrum of the Fokker-Planck equation (\ref{g2}) with diffusion coefficient (\ref{g4}).

\subsection{The relaxation time of a test vortex in a bath}
\label{sec_rbath}

The above derivation of the Fokker-Planck equation, relying on a {\it bath approximation}, assumes that the distribution of the field vortices is ``frozen'' so that their vorticity profile $\omega(r)$ does not evolve in time. This is always true for a thermal bath (\ref{tb1}), corresponding to a distribution  at statistical equilibrium (Boltzmann), because it does not evolve at all. For a stable steady state of the 2D Euler equation with a monotonic profile of angular velocity, this is true only on a timescale shorter than the relaxation time $t_R$ of the system as a whole. However, as we have indicated in Section \ref{sec_relaxwhole}, this timescale is very long because the relaxation time $t_R$ of the system as a whole  is larger than $Nt_D$.

Recalling that $\gamma\sim 1/N$, the Fokker-Planck operator in Eqs. (\ref{m4}) and (\ref{tb6}) scales like $\ln N/N$ (see Eq. (\ref{m6}) for the logarithmic correction). Therefore, the distribution $P(r,t)$ of the test vortex relaxes towards the distribution $\omega(r)$ of the bath on a typical time
\begin{equation}
t_{R}^{bath}\sim \frac{N}{\ln N}t_D.
\label{rbath1}
\end{equation}
This is the timescale controlling the relaxation of the test vortex, i.e. the time needed by the test vortex to acquire the distribution of the bath (see Appendix \ref{sec_reltime} for a more precise estimate). It should not be confused with the timescale (\ref{rw2}) controlling the relaxation of the system as a whole.  Since the timescale $t_{R}^{bath}$ is shorter than the timescale $t_R$ on which $\omega(r,t)$ changes due to collisions (in the case where $\omega(r,t)$ has a monotonic profile of angular velocity), we can consider that the distribution of the field vortices is ``frozen'' on the timescale (\ref{rbath1}). Therefore, our bath approximation is justified on this timescale and our approach is self-consistent.

\section{Conclusion}
\label{sec_conclusion}

In this paper, we have developed the kinetic theory of point vortices
by taking collective effects into account. We have used the formalism
of Dubin \& O'Neil \cite{dubin} valid for axisymmetric mean
flows. This improves our previous works
\cite{preR,pre,bbgky,copenhaguen,kindetail} where these effects were
neglected. Collective effects amount to replacing the ``bare''
potential of interaction by a ``dressed'' potential, without altering
the overall structure of the kinetic equation. In plasma physics,
collective effects included in the Lenard-Balescu
\cite{lenard,balescu} equation are important because they take into
account screening effects and regularize, at the scale of the Debye
length, the logarithmic divergence at large scales that appears in the
Landau \cite{landau} equation. In the case of point vortices, there is
no divergence in the Landau-type kinetic equation (\ref{lb29}) that
ignores collective effects, so that their influence may be less
important than in plasma physics. We have also developed a test vortex
approach and a Fokker-Planck theory. We have obtained the expressions of the
diffusion coefficient and drift term directly from the equations of
motion, taking collective effects into account. We have shown that
they can also be obtained from the kinetic equation by making a
bath approximation leading to a Fokker-Planck equation.  We have
presented the results for axisymmetric flows, but similar
results can be obtained for unidirectional flows \cite{preR,pre}.

The kinetic theory developed in this paper is valid at the order $1/N$ so that it describes the evolution of the system on a timescale $Nt_D$. This is sufficient to study the relaxation of a  test vortex in a bath of field vortices since the corresponding  relaxation time is of order $(N/\ln N)t_D$ (see Section \ref{sec_rbath}). However, this is not sufficient to describe the evolution of the system as a whole towards the statistical equilibrium state (Boltzmann) because, for axisymmetric flows, the relaxation time is larger than $Nt_D$ (see Section \ref{sec_relaxwhole}). Therefore, we need to develop the kinetic theory at higher orders (taking into account three-body, four-body,... correlations). At present, this is not done and the scaling of the relaxation time with $N$ remains an open problem.

We can try to make speculations by using analogies with other systems
with long-range interactions which present similar features
\cite{kindetail}. The most natural scaling would be $N^2 t_D$ which
corresponds to the next order term in the expansion of the basic
equations of the kinetic theory in powers of $1/N$
\cite{hb4,bbgky,kindetail}. An $N^2$ scaling is indeed obtained
numerically \cite{dawson,rouetfeix} for spatially homogeneous
one-dimensional plasmas for which the Lenard-Balescu collision term
vanishes at the order $1/N$ \cite{feix,kp}. However, the scaling of
the relaxation time may be more complex. For example, for the
permanently spatially homogeneous HMF model (for which the
Lenard-Balescu collision term also vanishes at the order $1/N$
\cite{bd,cvb}), Campa {\it et al.} \cite{campa} report a relaxation
time scaling like $e^N t_D$ (this timescale is, however, questioned in
recent works \cite{private} who find a $N^2t_D$ scaling).  It could
also happen that the point vortex gas never reaches statistical
equilibrium, i.e. the evolution may be non-ergodic \cite{khanin}. All
these speculations \cite{kindetail} could be checked numerically by
solving the $N$-vortex dynamics. The situation should be different for
more general flows that are not axisymmetric. In that case, there are
potentially more resonances so that the relaxation time could be
reduced and achieve the natural scaling $Nt_D$ corresponding to the
first order term in the kinetic theory
\cite{bbgky,kindetail}. Similarly, one dimensional spatially
homogeneous systems with long-range interaction are expected to relax
towards statistical equilibrium on a timescale $Nt_D$ due to
additional resonances \cite{angleaction}. An $N t_D$ scaling is indeed
obtained numerically for spatially inhomogeneous one dimensional
stellar systems
\cite{brucemiller,gouda,valageas,joyce}, for the spatially
inhomogeneous HMF model \cite{ruffoN} and for the relaxation of a
non-axisymmetric distribution of point vortices
\cite{kn}. On the other hand, for the HMF model, if an initially
spatially homogeneous distribution function becomes Vlasov unstable
during the collisional evolution, a dynamical phase transition from a
non-magnetized to a magnetized state takes place (as theoretically
studied in \cite{campachav}) and the relaxation time could be {\it
intermediate} between $N^2 t_D$ (permanently homogeneous) and $N t_D$
(permanently inhomogeneous). In that situation, Yamaguchi {\it et al.}
\cite{yamaguchi} find a relaxation time scaling like $N^{\delta}t_D$
with $\delta=1.7$. The previous argument (leading to $1<\delta<2$) may
provide a first step towards the explanation of this anomalous
exponent. The same phenomenon (loss of Euler stability due to
collisions and dynamical phase transition from an axisymmetric
distribution to a non-axisymmetric distribution) could happen for the
point vortex system.

Finally, we would like to conclude by briefly mentioning other works
on the kinetic theory of point vortices that are related to our study.
Schecter \& Dubin \cite{sdprl,sd2} have considered the motion of a
point vortex in a background vorticity gradient (without
fluctuation). They obtain an expression of the drift that coincides
with the one obtained previously in \cite{preR,pre}. However, the
point of view is different because they consider the evolution of a
test vortex in an external vorticity field without fluctuation, while
we consider the evolution of a test vortex in a bath of field vortices
that creates a vorticity gradient but also contains fluctuations. The
vorticity gradient leads to a (deterministic) drift as a result of a
polarization process and the fluctuations lead to a
diffusion. Therefore, the test vortex has a stochastic motion that can
be modeled by a Fokker-Planck equation of a drift-diffusion
type. There is no Fokker-Planck equation in the approach of Schecter
\& Dubin \cite{sdprl,sd2} since the evolution of the test vortex is
purely deterministic. On the other hand, Dubin \& Jin \cite{dubinjin}
consider the diffusion of point vortices in an external shear. They
obtain an expression of the diffusion coefficient that coincides with
the one obtained previously in \cite{preR,pre} from the Kubo formula
(in particular, they recover the shear reduction found in
\cite{preR}). However, the point of view is different because they
consider the effect of an external shear without vorticity
gradient. Therefore, there is no drift. By contrast, in our approach,
the test vortex moves in a bath of field vortices that presents a
vorticity gradient. In addition to its diffusive motion, the test
vortex modifies the distribution of the field vortices (as in a
polarization process) and the retroaction of the field vortices causes
the drift of the test vortex. In a sense, the works of Dubin and
collaborators isolate the effects of drift \cite{sdprl,sd2} and
diffusion \cite{dubinjin} while they are taken into account {\it
simultaneously} in our approaches
\cite{preR,pre,bbgky,copenhaguen,kindetail}. The extension of the
kinetic theory to the case of point vortices with different values of
individual circulation $\gamma$ has been performed in
\cite{marmanis,zakharov,dubin2,newtonmezic,clemou}. The statistics of
the velocity fluctuation arising from a random distribution of point
vortices has been studied in \cite{mezic,jimenez,chavsire1,chavsire2}
by analogy with the study of Chandrasekhar \& von Neumann \cite{cvn}
on the statistics of the gravitational force arising from a random
distribution of stars. This theory has been used in \cite{chavsire1} to
obtain an expression $D\propto \gamma (\ln N)^{1/2}$ of the diffusion
coefficient of point vortices in the absence of shear (i.e. in the
opposite limit to the one considered here). This result has been
applied to the problem of 2D decaying turbulence in
\cite{renormalization} in order to obtain the expression of the
exponent of anomalous diffusion $\langle x^2\rangle\sim t^{1+\xi/2}$
where $\xi$ is the exponent controlling the decay of the vortex number
$N\sim t^{-\xi}$ (it is argued in
\cite{renormalization,sirechavnew} that $\xi=1$ in the {\it strict} asymptotic
scaling regime $t\rightarrow +\infty$ so that $\nu\equiv 1+\xi/2=3/2$). The
energy spectrum of the point vortex gas has been determined in
\cite{novikov,chavsire2,houchesPH,ys,syyt}. Finally, a quasilinear
theory of the 2D Euler equation has been developed in
\cite{quasilinear} in order to describe a phase of ``gentle'' collisionless
relaxation. For general references about vortex dynamics, see
\cite{aref,boatto}.

 \appendix

\section{Potential of interaction}
\label{sec_pot}

We assume that the stream function $\psi({\bf r})$ is related to the vorticity $\omega({\bf r})$ by an equation of the form $L\psi=-\omega$,
where $L$ is a linear operator of the Laplacian type. The potential of interaction $u({\bf r},{\bf r}')$, which is the Green function of the operator $L$, is defined by $Lu({\bf r},{\bf r}')=-\delta({\bf r}-{\bf r}')$. For simplicity, we shall consider an infinite domain where $u=u(|{\bf r}-{\bf r}'|)$ only depends on the absolute distance between two points (our results can be extended to more general situations by using the Green functions of Lin \cite{lin}). In that case, the stream function is  related to the vorticity field by an expression of the form
\begin{equation}
\psi({\bf r},t)=\int u(|{\bf r}-{\bf r}'|)\omega({\bf r'},t)\, d{\bf r}.
\label{pot1}
\end{equation}
It can be written $\psi=u*\omega$ where $*$ denotes the product of convolution. For ordinary flows, the  stream function is related to the vorticity by the Poisson equation  $\Delta\psi=-\omega$. In that case $L=\Delta$ and the potential of interaction in an infinite domain is given by $u(|{\bf r}-{\bf r}'|)=-(1/2\pi)\ln|{\bf r}-{\bf r}'|$ which corresponds to a Newtonian (or Coulombian) interaction in two dimensions. In the quasigeostrophic (QG) model describing  geophysical flows \cite{pedlosky}, the  stream function is related to the (potential) vorticity by the screened Poisson equation  $\Delta\psi-k_R^2\psi=-\omega$ where $k_R^{-1}$ is the Rossby radius. In that case $L=\Delta-k_R^2$ and the potential of interaction in an infinite domain is given by $u(|{\bf r}-{\bf r}'|)=(1/2\pi)K_0(k_R|{\bf r}-{\bf r}'|)$. The previous results are recovered for $k_R\rightarrow 0$.

Let us first consider the ordinary potential corresponding to $L=\Delta$. Introducing polar coordinates, the Poisson equation can be written
\begin{equation}
\frac{1}{r}\frac{\partial}{\partial r} r \frac{\partial \delta\psi}{\partial r}+\frac{1}{r^2}\frac{\partial^2\delta\psi}{\partial\theta^2}=-\delta\omega.
\label{pot2}
\end{equation}
Taking the Fourier-Laplace transform of this equation,  we obtain
\begin{equation}
\left\lbrack \frac{1}{r}\frac{\partial}{\partial r} r \frac{\partial}{\partial r}-\frac{n^2}{r^{2}}\right \rbrack \delta\tilde\psi(n,r,\sigma)=-\delta\tilde\omega(n,r,\sigma).
\label{pot3}
\end{equation}
This is of the form ${\cal L}\delta\tilde\psi=-\delta\tilde\omega$ with ${\cal L}=\frac{1}{r}\frac{\partial}{\partial r} r \frac{\partial}{\partial r}-\frac{n^2}{r^{2}}$. Substituting Eq. (\ref{lb9}) in Eq. (\ref{pot3}), we find that
\begin{equation}
\left\lbrack \frac{1}{r}\frac{\partial}{\partial r} r \frac{\partial}{\partial r}-\frac{n^2}{r^{2}}-\frac{n\frac{1}{r}\frac{\partial\omega}{\partial r}}{n\Omega-\sigma}\right \rbrack \delta\tilde\psi(n,r,\sigma)=-\frac{\delta\hat\omega(n,r,0)}{i(n\Omega-\sigma)}.
\label{pot4}
\end{equation}
Therefore $\delta\tilde\psi(n,r,\sigma)$ is related to $\delta\hat\omega(n,r,0)$ by a relation of the form (\ref{lb11}) where the Green function $G(n,r,r',\sigma)$ is defined by
\begin{equation}
\left\lbrack \frac{1}{r}\frac{\partial}{\partial r} r \frac{\partial}{\partial r}-\frac{n^2}{r^{2}}-\frac{n\frac{1}{r}\frac{\partial\omega}{\partial r}}{n\Omega-\sigma}\right \rbrack G(n,r,r',\sigma)=-\frac{\delta(r-r')}{2\pi r}.
\label{pot5}
\end{equation}
If we neglect collective effects in the  foregoing equation, we obtain
\begin{equation}
\left\lbrack \frac{1}{r}\frac{\partial}{\partial r} r \frac{\partial}{\partial r}-\frac{n^2}{r^{2}}\right \rbrack G_{bare}(n,r,r')=-\frac{\delta(r-r')}{2\pi r}.
\label{pot6}
\end{equation}
This shows that the bare Green function $G_{bare}(n,r,r')=\hat{u}_n(r,r')$ is the Fourier transform of the potential of interaction $u$ that is solution of the Poisson equation $\Delta u=-\delta$.

Let us now consider a general form of interaction between point vortices. The fluctuations of the stream function are related to the fluctuations of the vorticity by
\begin{equation}
\delta\psi({\bf r},t)=\int u(|{\bf r}-{\bf r}'|)\delta\omega({\bf r'},t)\, d{\bf r}.
\label{pot7}
\end{equation}
The potential of interaction can be written
\begin{equation}
u(|{\bf r}-{\bf r}'|)=u\left (\sqrt{r^2+r'^2-2rr'\cos(\theta-\theta')}\right )=u(r,r',\phi),
\label{pot8}
\end{equation}
where $\phi=\theta-\theta'$. Due to its $\phi$-periodicity, it can be decomposed in Fourier series of the form
\begin{equation}
u(r,r',\phi)=\sum_n e^{i n \phi}\hat{u}_n(r,r'),\qquad \hat{u}_n(r,r')=\int_{0}^{2\pi}\frac{d\phi}{2\pi}u(r,r',\phi)\cos(n\phi).
\label{pot9}
\end{equation}
Taking the Fourier-Laplace transform of Eq. (\ref{pot7}) and using the fact that the integral is a product of convolution, we get
\begin{equation}
\delta\tilde\psi(n,r,\sigma)=2\pi \int_0^{+\infty} r'dr'\,  \hat{u}_n(r,r') \delta\tilde\omega(n,r',\sigma).
\label{pot10}
\end{equation}
If we introduce Eq. (\ref{lb9}) in Eq. (\ref{pot10}), we obtain a Fredholm integral equation
\begin{equation}
\delta\tilde\psi(n,r,\sigma)+2\pi \int_0^{+\infty} r'dr'\,  \hat{u}_n(r,r')\frac{n\frac{1}{r'}\frac{\partial\omega'}{\partial r'}}{n\Omega'-\sigma}\delta\tilde\psi(n,r',\sigma) =2\pi \int_0^{+\infty} r'dr'\,  \hat{u}_n(r,r') \frac{\delta\hat\omega(n,r',0)}{i(n\Omega'-\sigma)},
\label{pot11}
\end{equation}
which is equivalent to Eq. (\ref{lb10}). If we neglect collective effects, the foregoing equation reduces to
\begin{equation}
\delta\tilde\psi(n,r,\sigma) =2\pi \int_0^{+\infty} r'dr'\,  \hat{u}_n(r,r') \frac{\delta\hat\omega(n,r,0)}{i(n\Omega-\sigma)}.
\label{pot12}
\end{equation}
Comparing Eq. (\ref{pot12}) with Eq. (\ref{lb11}), we see that $G_{bare}(n,r,r')=\hat{u}_n(r,r')$ is just the Fourier transform of the ``bare'' potential of interaction $u$. This is equivalent to ${\cal L}G_{bare}(n,r,r')=-\delta(r-r')/2\pi r$. When collective effects are taken into account, Eq. (\ref{pot12}) is replaced by Eq. (\ref{lb11}) where the ``dressed'' potential of interaction satisfies
\begin{equation}
G(n,r,r',\sigma)+2\pi \int_0^{+\infty} r''dr''\,  \hat{u}_n(r,r'')\frac{n\frac{1}{r''}\frac{\partial\omega''}{\partial r''}}{n\Omega''-\sigma}G(n,r'',r',\sigma) =\hat{u}_n(r,r'),
\label{pot11ht}
\end{equation}
which is equivalent to Eq. (\ref{lb12}). 

For the Newtonian interaction $u(|{\bf r}-{\bf r}'|)=-(1/2\pi)\ln|{\bf r}-{\bf r}'|$, we have $u(r,r',\phi)=-(1/4\pi)\ln(r^2+r'^2-2rr'\cos\phi)$. The integrals in Eq. (\ref{pot9}) can be performed analytically \cite{kindetail}, and we obtain
\begin{equation}
\hat{u}_n(r,r')=\frac{1}{4\pi |n|}\left (\frac{r_<}{r_>}\right )^{|n|},\qquad \hat{u}_0(r,r')=-\frac{1}{2\pi}\ln r_>.
\label{pot13}
\end{equation}
Therefore
\begin{equation}
u(r,r',\phi)=-\frac{1}{2\pi}\ln r_>+\frac{1}{4\pi}\sum_{n\neq 0}\frac{1}{|n|}\left (\frac{r_<}{r_>}\right )^{|n|}e^{in\phi},
\label{pot14}
\end{equation}
which is the Fourier decomposition of the logarithm in two dimensions. The function defined by Eq. (\ref{lb31}) can be written \cite{pre}:
\begin{equation}
\chi_{bare}(r,r')=\frac{1}{8\pi^2}\sum_{n=1}^{+\infty}\frac{1}{n}\left (\frac{r_<}{r_>}\right )^{2n}=-\frac{1}{8\pi^2}\ln\left\lbrack 1-\left (\frac{r_<}{r_>}\right )^2\right\rbrack.
\label{pot15}
\end{equation}
Taking $r'=r$ in the foregoing expression, we obtain
\begin{equation}
\chi_{bare}(r,r)=\frac{1}{8\pi^2}\sum_{n=1}^{+\infty}\frac{1}{n}=\frac{1}{8\pi^2}\ln\Lambda,
\label{pot16}
\end{equation}
where $\ln\Lambda$ is the Coulomb factor (see Sec. \ref{sec_m}). Analogous expressions, valid in a circular domain of size $R$, are given in \cite{bbgky}.

\section{Auto-correlation of the fluctuations of the one-particle density}
\label{sec_auto}

According to Eq. (\ref{lb8}), we have
\begin{eqnarray}
\langle \delta\hat\omega(n,r,0)\delta\hat\omega(n',r',0)\rangle&=&\int_0^{2\pi}\frac{d\theta}{2\pi}\int_0^{2\pi}
\frac{d\theta'}{2\pi}e^{-i(n\theta+n'\theta')}\langle \delta\omega(\theta,r,0)\delta\omega(\theta',r',0)\rangle\nonumber\\
&=&\int_0^{2\pi}\frac{d\theta}{2\pi}\int_0^{2\pi}
\frac{d\theta'}{2\pi}e^{-i(n\theta+n'\theta')}\left\lbrack \langle \omega_d(\theta,r,0)\omega_d(\theta',r',0)\rangle-\omega(r)\omega(r')\rangle\right\rbrack.
\label{auto1}
\end{eqnarray}
The expression (\ref{pvm4}) of the discrete vorticity distribution leads to
\begin{eqnarray}
\langle \omega_d(\theta,r,0)\omega_d(\theta',r',0)\rangle&=&\gamma^2\sum_{i,j}\left\langle \delta(\theta-\theta_i)\frac{\delta(r-r_i)}{r}\delta(\theta'-\theta_j)\frac{\delta(r'-r_j)}{r'}\right\rangle\nonumber\\
&=&\gamma^2\sum_{i}\left\langle \delta(\theta-\theta_i)\frac{\delta(r-r_i)}{ r}\delta(\theta-\theta')\frac{\delta(r-r')}{r}\right\rangle\nonumber\\
&+&\gamma^2\sum_{i\neq j}\left\langle \delta(\theta-\theta_i)\frac{\delta(r-r_i)}{r}\delta(\theta'-\theta_j)\frac{\delta(r'-r_j)}{r'}\right\rangle\nonumber\\
&=&2\pi\gamma \omega(r) \delta(\theta-\theta')\frac{\delta(r-r')}{2\pi r}+\omega(r)\omega(r'),
\label{auto2}
\end{eqnarray}
where we have assumed that there is no correlation initially (if there are initial correlations, it can be shown that they are washed out rapidly \cite{pitaevskii,cdr} so that they have no effect on the final form of the collision term). Combining Eqs. (\ref{auto1}) and (\ref{auto2}), we obtain
\begin{equation}
\langle \delta\hat\omega(n,r,0)\delta\hat\omega(n',r',0)\rangle=\gamma\delta_{n,-n'}\frac{\delta(r-r')}{2\pi r}\omega(r).
\label{auto3}
\end{equation}

\section{The function $K(r,r',t)$}
\label{sec_k}

The kinetic equation (\ref{lb30}) can be rewritten in the more compact form
\begin{equation}
\frac{\partial\omega}{\partial t}=\frac{1}{r}\frac{\partial}{\partial r}r\int_{0}^{+\infty}  dr'  \, K(r,r',t) \left (\frac{1}{r}\frac{\partial}{\partial r}-\frac{1}{r'}\frac{\partial}{\partial r'}\right )\omega(r,t)\omega(r',t),
\label{k1}
\end{equation}
with
\begin{equation}
K(r,r',t)= 2\pi^2 \gamma \frac{r'}{r} \chi(r,r',t) \delta(\Omega(r,t)-\Omega(r',t)).
\label{k2}
\end{equation}
In the thermal bath approximation, the diffusion coefficient and the drift terms can  be rewritten
\begin{equation}
D(r)=\frac{1}{r}\int_{0}^{+\infty} dr' \, K(r,r') \omega(r'),\quad V_{drift}(r)=\int_{0}^{+\infty} \frac{dr'}{r'} \, K(r,r') \frac{d \omega}{d r}(r'),
\label{k3}
\end{equation}
\begin{equation}
A(r)=\int_{0}^{+\infty} r r' dr' \, \omega(r')\left (\frac{1}{r}\frac{\partial}{\partial r}-\frac{1}{r'}\frac{\partial}{\partial r'}\right )\frac{K(r,r')}{rr'}.
\label{k5}
\end{equation}

\section{A more precise estimate of the relaxation time}
\label{sec_reltime}

When the field vortices have the Gaussian distribution (\ref{g1}), a
more precise estimate of the relaxation time can be given. We first
define the ``vortex size'' $R$ by $R^2=\langle r^2\rangle=L/\Gamma$
where $L$ is the angular momentum. For a Gaussian distribution, $R^2=2/\lambda\gamma$. 
If we set $x=\sqrt{\lambda\gamma/2}r$, the Fokker-Planck equation
(\ref{g2}) can be rewritten
\begin{equation}
\label{r1}{\partial P\over\partial
t}=\frac{1}{t_R}\frac{1}{x}{\partial\over\partial x}\biggl\lbrack x G(x)\biggl
({\partial P\over\partial x}+2 Px\biggr
)\biggr\rbrack,
\end{equation}
with
\begin{equation}
\label{r2}
G(x)=\frac{x^2}{e^{x^2}-x^2-1}, \qquad t_R=\frac{8}{\lambda\gamma^2\ln\Lambda}.
\end{equation}
The ``reference time'' $t_R$ gives an estimate of the relaxation time of the test vortex in a Gaussian bath. It we define the dynamical time by $t_D=R^2/\Gamma=2/\lambda\gamma\Gamma$, we obtain
\begin{equation}
\label{r3}
t_R=\frac{4\Gamma}{\gamma\ln\Lambda}\sim \frac{8N}{\ln N}t_D,
\end{equation}
where we have used Eq. (\ref{m6}) to get the equivalent for $N\rightarrow +\infty$. We can also estimate the relaxation time as follows. The typical position of the test vortex increases like $\langle r^2\rangle\sim 4D(R)t$. The relaxation time $t_r$ is the time needed by the vortex to diffuse over a distance $R$. Taking $\langle r^2\rangle=R^2$ in the foregoing formula, we obtain
\begin{equation}
\label{r4}
t_r=\frac{R^2}{4D(R)}= \frac{2}{\lambda\gamma^2G(1)\ln\Lambda}=0.18t_R,
\end{equation}
where we have used $G(1)=1/(e-2)\simeq 1.39221$. Finally, although the  relaxation towards the Gaussian distribution is not exponential, a measure of the ``relaxation time'' is provided by  $t'_r=\xi^{-1}$ where $\xi=D(R)\lambda\gamma$ is the drift coefficient (mobility) given by an Einstein formula. This yields
\begin{equation}
\label{r5}
t_r'=\frac{1}{D(R)\lambda\gamma}= \frac{4}{\lambda\gamma^2G(1)\ln\Lambda}=2t_r.
\end{equation}
The same arguments can be extended to the Fokker-Planck equations (\ref{m4}) and (\ref{tb6}), leading to the scaling (\ref{rbath1}) of the relaxation time, up to a numerical factor.


\begin{thebibliography}{}

\bibitem{newton}  {\small P.K. Newton, The $N$-Vortex Problem: Analytical Techniques, in: Applied Mathematical Sciences, vol. 145, Springer-Verlag, Berlin, 2001 }
\bibitem{houches}  {\small  {\it Dynamics and Thermodynamics of Systems with Long-Range Interactions}, edited by T. Dauxois, S. Ruffo, E. Arimondo and  M. Wilkens, Lectures  Notes in Physics {\bf 602} (Berlin: Springer, 2002)}
\bibitem{assise}  {\small {\it Dynamics and Thermodynamics of Systems with Long-Range
Interactions: Theory and Experiments}, edited by A. Campa, A. Giansanti, G. Morigi and F. Sylos Labini, AIP Conf. Proc. {\bf 965} 122 (2008)}
\bibitem{oxford}  {\small  {\it Long-Range Interacting Systems}, edited by T. Dauxois, S. Ruffo and L. Cugliandolo, Les Houches Summer School 2008, (Oxford: Oxford University Press, 2009)}
\bibitem{cdr}  {\small A. Campa, T. Dauxois, S. Ruffo,   Physics Reports {\bf 480}, 57 (2009)}
\bibitem{onsager}  {\small  L. Onsager,   Nuovo Cimento, Suppl. {\bf  6}, 279 (1949)  }
\bibitem{purcell} {\small  E.M. Purcell, R.V. Pound, Phys. Rev. {\bf 81}, 279 (1951)}
\bibitem{jm} {\small  G. Joyce, D. Montgomery, J. Plasma Phys.  {\bf 10}, 107 (1973)}
\bibitem{mj}  {\small  D. Montgomery, G. Joyce,  Phys. Fluids {\bf 17}, 1139  (1974)}
\bibitem{kida}  {\small  S. Kida,  J. Phys. Soc. Jpn.  {\bf 39}, 1395  (1975)}
\bibitem{pl}  {\small  Y.B. Pointin, T.S. Lundgren,  Phys. Fluids. {\bf  19}, 1459 (1976)  }
\bibitem{lp}  {\small  T.S. Lundgren, Y.B. Pointin, J. Stat. Phys. {\bf  17}, 323 (1977)  }
\bibitem{esree}  {\small G.L. Eyink, K.R. Sreenivasan, Rev. Mod. Phys. {\bf 78}, 87 (2006)}
\bibitem{caglioti}  {\small E. Caglioti, P.L. Lions, C. Marchioro, M.  Pulvirenti,  Commun. Math. Phys.   {\bf 143}, 501 (1992)   }
\bibitem{k93}  {\small M. Kiessling,  Commun. Pure Appl. Math.   {\bf 47}, 27 (1993)   }
\bibitem{es}  {\small G.L. Eyink, H. Spohn,   J. Stat. Phys.  {\bf 70}, 833 (1993)  }
\bibitem{ca2}  {\small  E. Caglioti, P.L. Lions, C.  Marchioro, M.  Pulvirenti,  Commun. Math. Phys.   {\bf 174}, 229 (1995)}
\bibitem{kl}  {\small M. Kiessling, J.  Lebowitz, Lett. Math. Phys.   {\bf 42}, 43 (1997)}
\bibitem{khanin}  {\small K.M. Khanin,  Physica D {\bf 2}, 261 (1982)}
\bibitem{dubin}  {\small D. Dubin, T.M. O'Neil,  Phys. Rev. Lett. {\bf 60}, 1286 (1988)}
\bibitem{lenard}  {\small A. Lenard, Ann. Phys. (N.Y.) {\bf 10}, 390 (1960)}
\bibitem{balescu}  {\small   R. Balescu, Phys. Fluids {\bf 3}, 52 (1960)}
\bibitem{sdprl} {\small D.A. Schecter, D. Dubin, Phys. Rev. Lett. {\bf 83}, 2191 (1999)}
\bibitem{sd2}  {\small D.A. Schecter, D. Dubin, Phys. Rev. E {\bf 13}, 1704 (2001)}
\bibitem{dubinjin}  {\small D. Dubin, D.Z. Jin, Phys. Lett. A {\bf 284}, 112 (2001)}
\bibitem{dubin2}  {\small D. Dubin, Phys. Plasmas {\bf 10}, 1338 (2003)}
\bibitem{preR}  {\small P.H. Chavanis, Phys. Rev. E {\bf 58}, R1199 (1998)}
\bibitem{pre}  {\small P.H. Chavanis, Phys. Rev. E {\bf 64}, 026309 (2001)}
\bibitem{bbgky}  {\small  P.H. Chavanis,  Physica A {\bf 387}, 1123 (2008)  }
\bibitem{copenhaguen}  {\small P.H. Chavanis, Theor. Comput. Fluid Dyn.  {\bf  24}, 217 (2010)}
\bibitem{kindetail} {\small P.H. Chavanis, J. Stat. Mech. (2010) P05019}
\bibitem{chandra}  {\small S. Chandrasekhar, Principles of Stellar Dynamics (University of Chicago press, 1942)}
\bibitem{chandra1}  {\small S. Chandrasekhar, Astrophys. J.  {\bf 97}, 255 (1943) }
\bibitem{nice}  {\small S. Chandrasekhar, Rev. Mod. Phys. {\bf 21}, 383 (1949) }
\bibitem{kandrup1}  {\small H. Kandrup,  Astrophys. J. {\bf 244}, 316 (1981)}
\bibitem{kandrup2}  {\small H. Kandrup,   Astro. Space. Sci.  {\bf 97}, 435 (1983)}
\bibitem{landau}  {\small L.D. Landau, Phys. Z. Sowj. Union  {\bf 10}, 154 (1936)}
\bibitem{clemou}  {\small  P.H. Chavanis, M. Lemou, Eur. Phys. J. B {\bf 59}, 217 (2007)}
\bibitem{sano}  {\small M.M. Sano,  Phys. Rev. E {\bf 76}, 046312 (2007)}
\bibitem{pedlosky} {\small J. Pedlosky, Geophysical Fluid Dynamics (Springer, Berlin, 1987)}
\bibitem{csr}  {\small  P.H. Chavanis, J. Sommeria, R.  Robert, Astrophys. J.  {\bf 471}, 385 (1996)}
\bibitem{houchesPH}  {\small P.H. Chavanis,  {\it Statistical mechanics of two-dimensional vortices and stellar systems}, in:  Dynamics and thermodynamics of systems with long range interactions, edited by Dauxois, T, Ruffo, S., Arimondo, E. and  Wilkens, M. Lecture Notes in Physics, Springer (2002)}
\bibitem{ar}  {\small  M. Antoni, S.  Ruffo,  Phys. Rev. E {\bf  52}, 2361 (1995)  }
\bibitem{inagakikin}  {\small S. Inagaki, Prog. Theor. Phys.  {\bf 96}, 1307 (1996)}
\bibitem{bouchet}  {\small F. Bouchet, Phys. Rev. E {\bf 70}, 036113 (2004) }
\bibitem{bd}  {\small F. Bouchet, T.  Dauxois,  Phys. Rev. E {\bf 72}, 045103 (2005)}
\bibitem{cvb}  {\small  P.H. Chavanis, J. Vatteville, F.  Bouchet,  Eur. Phys. J. B {\bf  46}, 61 (2005)}
\bibitem{clpre}  {\small P.H. Chavanis, M.  Lemou,  Phys. Rev. E {\bf 72}, 061106  (2005) }
\bibitem{curious}  {\small  P.H. Chavanis,  Eur. Phys. J. B {\bf  52}, 47 (2006)}
\bibitem{bgm} {\small F. Bouchet, S. Gupta, D. Mukamel, Physica A {\bf 389}, 4389 (2010)}
\bibitem{feix}  {\small O.C. Eldridge, M. Feix,  Phys. Fluids.  {\bf 6}, 398 (1962) }
\bibitem{kp}  {\small B.B. Kadomtsev, O.P. Pogutse, Phys. Rev. Lett.  {\bf 25}, 1155 (1970) }
\bibitem{angleaction}  {\small  P.H. Chavanis,  Physica A {\bf 377}, 469 (2007)  }
\bibitem{hubbard}  {\small J. Hubbard, Proc. R. Soc. Lond. {\bf 260}, 114 (1961) }
\bibitem{kirchhoff}  {\small  G. Kirchhoff, in {\it Lectures in Mathematical Physics, Mechanics} (Teubner, Leipzig, 1877). }
\bibitem{pitaevskii}  {\small  E.M. Lifshitz, L.P.  Pitaevskii,
{Physical Kinetics} (Pergamon Press, Oxford, 1981)}
\bibitem{bt}  {\small  J. Binney, S.  Tremaine, {Galactic Dynamics} (Princeton Series in Astrophysics, 1987)}
\bibitem{chen}  {\small  P. Chen, M.C. Cross,   Phys. Rev. Lett.  {\bf 77}, 4174 (1996)}
\bibitem{brands}  {\small  H. Brands, P.H. Chavanis, R. Pasmanter, J. Sommeria, Phys. Fluids {\bf 11}, 3465 (1999)}
\bibitem{miller}  {\small  J. Miller,   Phys. Rev. Lett.  {\bf 65}, 2137 (1990)}
\bibitem{rs}  {\small  R. Robert, J. Sommeria, J. Fluid Mech.  {\bf 229}, 291 (1991)}
\bibitem{lb}  {\small  D. Lynden-Bell, Mon. Not. R. Astron. Soc.  {\bf 136}, 101 (1967)}
\bibitem{joyceLB}{\small M. Joyce, T. Worrakitpoonpon, [arXiv1012.5042]}
\bibitem{hb4}  {\small  P.H. Chavanis,  Physica A  {\bf 387}, 1504 (2008)}
\bibitem{boghosian}  {\small  B.M. Boghosian,  Phys. Rev. E  {\bf 53}, 4754 (1996)}
\bibitem{chavcampa}  {\small  P.H. Chavanis, A. Campa, Eur.  Phys. J. B  {\bf 76}, 581 (2010)}
\bibitem{paddy}  {\small  T. Padmanabhan, Phys. Rep. {\bf 188}, 285 (1990)}
\bibitem{hb2}  {\small P.H. Chavanis, Physica A  {\bf 361}, 81 (2006)}
\bibitem{kawahara}  {\small R. Kawahara, H. Nakanishi,  J. Phys. Soc. Japan {\bf 76}, 074001 (2007)}
\bibitem{yoshida}  {\small T. Yoshida,  J. Phys. Soc. Japan {\bf 78}, 024004 (2009)}
\bibitem{kn}  {\small R. Kawahara, H. Nakanishi,  J. Phys. Soc. Japan {\bf 75}, 054001 (2006)}
\bibitem{risken}  {\small  H. Risken
{The Fokker-Planck equation} (Springer, 1989)}
\bibitem{chavsire1}  {\small P.H. Chavanis, C. Sire, Phys. Rev. E {\bf 62}, 490 (2000)}
\bibitem{marksteiner}  {\small S. Marksteiner, K. Ellinger and P. Zoller, Phys. Rev. A {\bf 53}, 34 (1996)}
\bibitem{dawson}  {\small J. Dawson,  Phys. Fluids  {\bf 7}, 419 (1964)}
\bibitem{rouetfeix}  {\small J.L. Rouet, M. Feix,  Phys. Fluids B  {\bf 3}, 1830 (1991)}
\bibitem{campa}  {\small A. Campa, A.  Giansanti, G.  Morelli, Phys. Rev. E {\bf 76}, 041117 (2007) }
\bibitem{private}  {\small S. Ruffo, S. Gupta, private communication }
\bibitem{brucemiller} {\small B.N. Miller, Phys. Rev. E {\bf 53}, R4279 (1996)}
\bibitem{gouda} {\small T. Tsuchiya, N. Gouda, T. Konishi, Phys. Rev. E {\bf 53}, 2210 (1996)}
\bibitem{valageas}  {\small P. Valageas, Phys. Rev. E {\bf 74}, 016606 (2006) }
\bibitem{joyce} {\small M. Joyce, T. Worrakitpoonpon, J. Stat. Mech. {\bf 10}, 12 (2010)}
\bibitem{ruffoN} {\small P. de Buyl, D. Mukamel, S. Ruffo, [arXiv:1012.2594]}
\bibitem{campachav}  {\small A. Campa, P.H. Chavanis, A.  Giansanti, G. Morelli,  Phys. Rev. E {\bf 78}, 040102 (2008) }
\bibitem{yamaguchi}  {\small Y. Yamaguchi, J.   Barr\'e, F. Bouchet, T.  Dauxois, S. Ruffo,   Physica A {\bf 337}, 36 (2004)}
\bibitem{marmanis}  {\small H. Marmanis, Proc. R. Soc. Lond. A {\bf 454}, 587 (1998) }
\bibitem{zakharov}  {\small  S. Nazarenko, V.E. Zakharov, Physica D  {\bf  56}, 381 (1992)}
\bibitem{newtonmezic}  {\small P.K. Newton, I.  Mezic,  Journal of Turbulence {\bf 3}, 52 (2002)}
\bibitem{mezic} {\small I.A. Min, I. Mezic, A. Leonard, Phys. Fluids {\bf 8}, 1169 (1996)}
\bibitem{jimenez} {\small J. Jim\'enez, J. Fluid Mech. {\bf 313}, 223 (1996)}
\bibitem{chavsire2}  {\small P.H. Chavanis, C. Sire,  Phys. Fluids {\bf 13}, 1904 (2001)}
\bibitem{cvn}  {\small S. Chandrasekhar, J. von Neumann,  Astrophys. J. {\bf 95}, 489 (1942)}
\bibitem{renormalization}  {\small C. Sire, P.H.  Chavanis,  Phys. Rev. E  {\bf 61}, 6644 (2000)}
\bibitem{sirechavnew}  {\small C. Sire, P.H.  Chavanis, J. Sopik
[arXiv1006.3206]}
\bibitem{novikov}  {\small E.A. Novikov, Sov. Phys. JETP {\bf 41}, 937 (1975)}
\bibitem{ys}  {\small T. Yoshida, M.S. Sano, J. Phys. Soc. Jpn. {\bf 74}, 587 (2005)}
\bibitem{syyt}  {\small M.S. Sano, Y. Yatsuyanagi, T. Yoshida, H. Tomita, J. Phys. Soc. Jpn. {\bf 76}, 064001 (2007)}
\bibitem{quasilinear}  {\small  P.H. Chavanis,  Phys. Rev. Lett.  {\bf 84}, 5512 (2000)}
\bibitem{lin}  {\small C.C. Lin, Proc. Natl. Acad. Sci. {\bf 27}, 570 (1941)}
\bibitem{aref}  {\small {\it 150 Years of Vortex Dynamics}, edited by H. Aref, Proceedings of an IUTAM Symposium held at the Technical University of Denmark (Springer, 2008)}
\bibitem{boatto}  {\small S. Boatto, D. Crowdy, {\it Point vortex dynamics}, Encyclopedia of Mathematical Physics (Elsevier, 2006)}

	




\end{thebibliography}
\end{document}